*Review*

# From Random to Regular:

# Variation in the Patterning of Retinal Mosaics[*]

Patrick W. Keeley[1], Stephen J. Eglen[2] and Benjamin E. Reese[1,3] [†]

[1]Neuroscience Research Institute, University of California at Santa Barbara,

[2]Department of Applied Mathematics and Theoretical Physics, University of Cambridge,

and [3]Department of Psychological and Brain Sciences, University of California at Santa Barbara

*Key words:* Nearest Neighbor, Voronoi Domain, Regularity Index, Auto-correlation, Density Recovery Profile, Effective Radius, Packing Factor

*Running title:* From Random to Regular

*Acknowledgements:* This research was supported by a grant from the NIH (EY-019968)

*In memory of Lucia Galli-Resta, who advanced the application of spatial statistical analyses in the study of retinal mosaics

[†]Address correspondence to B.E. Reese, Neuroscience Research Institute, University of California at Santa Barbara, Santa Barbara CA 93106-5060; tel. 805-893-2091; breese@psych.ucsb.edu

**Abstract**

The various types of retinal neurons are each positioned at their respective depths within the retina where they are believed to be assembled as orderly mosaics, in which like-type neurons minimize proximity to one another.  Two common statistical analyses for assessing the spatial properties of retinal mosaics include the nearest neighbor analysis, from which an index of their "regularity" is commonly calculated, and the density recovery profile derived from auto-correlation analysis, revealing the presence of an exclusion zone indicative of anti-clustering.  While each of the spatial statistics derived from these analyses, the regularity index and the effective radius, can be useful in characterizing such properties of orderly retinal mosaics, they are rarely sufficient for conveying the natural variation in the self-spacing behavior of different types of retinal neurons and the extent to which that behavior generates uniform intercellular spacing across the mosaic.  We consider the strengths and limitations of different spatial statistical analyses for assessing the patterning in retinal mosaics, highlighting a number of misconceptions and their frequent misuse.  Rather than being diagnostic criteria for determining simply whether a population is "regular", they should be treated as descriptive statistics that convey variation in the factors that influence neuronal positioning.  We subsequently apply multiple spatial statistics to the analysis of eight different mosaics in the mouse retina, demonstrating conspicuous variability in the degree of patterning present, from essentially random to notably regular.  This variability in patterning has both a developmental as well as a functional significance, reflecting the rules governing the positioning of different types of neurons as the architecture of the retina is assembled, and the distinct mechanisms by which they regulate dendritic growth to generate their characteristic coverage and connectivity.

**Introduction**

The mammalian retina is composed of perhaps as many as a hundred different types of neurons (Zeng and Sanes, 2017). Each of those retinal cell types exhibits a unique transcriptional profile and morphological signature (Macosko et al., 2015; Shekhar et al., 2016; Sumbul et al., 2014), residing at a characteristic depth within the retina in one of three different cellular layers. These neuronal populations differ in other demographic properties as well, including the absolute numbers of their constituent cells, in how those cells are positioned relative to one another within their respective strata, and in the degree to which they spread their processes to overlap one another (Reese and Keeley, 2016), each of them uniquely integrated into the retinal circuitry to contribute to visual processing (Baden et al., 2016). The latter features are attributes of the "mosaic properties" of these populations, being histotypical features that are thought to ensure uniform service in retinal function across the visual field (Reese and Keeley, 2015).

The presumption that retinal neurons are assembled as orderly retinal mosaics is so widespread that "regularity" has come to be regarded as a cardinal, defining, feature of a cell type (Cook, 1998, 2003). It is true that some neuronal types exhibit an obvious patterning in their distributions, and these orderly arrays of well-spaced neurons can often be simulated by local spacing rules that preclude proximity between neighboring cells {Eglen, 2006 #1554; Eglen, 2003a #1090; Galli-Resta, 2001 #1262}; these mosaics are clearly "regular". Yet what exactly in a retinal mosaic defines its "regularity" is rarely examined, as different studies use different measures to ascribe this attribute to a labeled population of neurons. Two spatial statistics associated with these mosaics, the "regularity index" derived from an analysis of nearest neighbor distances and the "effective radius" determined from the density recovery profile (informally termed the exclusion zone), have each come to be regarded as definitive evidence that a population of labeled neurons is indeed "regular" due to "self-spacing" behavior, and consequently a distinct type of neuron. In fact, the bar is often set exceedingly low for identifying order in an array of neurons using either statistic. The present review will first consider the use of these different spatial statistics, demonstrating their limitations and occasional misuse in the analysis of retinal mosaics. We will subsequently show, through the examination of multiple populations of retinal neurons in the mouse retina, that there is conspicuous variability in the patterning within their mosaics, from essentially random to highly regular[1]. That variability reflects the cell-type specific regulation of neuronal positioning during the assembly of the retinal architecture, which, in conjunction with distinct mechanisms governing dendritic outgrowth, establish the coverage and connectivity characteristic of each cell type, a topic we consider elsewhere (Reese and Keeley, 2015).

***Nearest neighbor analysis and the regularity index***

The widespread adoption of the nearest neighbor analysis followed demonstrations by Wässle and coworkers that various low-density retinal mosaics like those of horizontal cells or retinal ganglion cells, detected by virtue of their selective neurofibrillar silver staining or through the retrograde labeling of their dendritic arbors, exhibit a patterning quite distinctive from random distributions of infinitesimally small (i.e. dimensionless) points (Wässle et al., 1981b, c; Wässle and Riemann, 1978). The

---

[1] The term "mosaic" will be used herein to refer to any population of cells sharing some attribute, for instance, being immuno-positive at a specific depth within the retina, thought to identify members of a particular cell type. Whether it is, in fact, a singular type of neuron is an empirical matter, but the basis for such a conclusion should be made independent of its patterning, be that defined in terms of its regularity or its minimal self-spacing. Our use of the term carries no presumption of a degree of orderliness within the array.

Gaussian nature of the distribution of the nearest neighbor distances derived from such mosaics contrasted with the theoretical Poisson distribution associated with random point patterns (e.g. figure 1a-d). By expressing the variation in those nearest neighbor distances relative to their average (specifically, the ratio of the mean nearest neighbor distance to the standard deviation), a summary statistic, the "regularity index", seemed to yield a simple quantitation of this property, increasing towards infinity as a mosaic approached a perfect lattice. Six particularly well-labeled samples were so analyzed in the cat's retina, with A-type horizontal cells having indices of 4.7 to 6.2 (being derived from two samples at different eccentricities) (Wässle and Riemann, 1978), ON and OFF alpha retinal ganglion cells having indices of ~4.6 (Wässle et al., 1981c), and ON and OFF beta retinal ganglion cells having indices of 5.3 (Wässle et al., 1981a), the latter pairs being derived from two fields in which the ON versus OFF cells of each type could be reliably distinguished. Those early studies focused upon the novel recognition that such patterning is present and that it might be assessed quantitatively, stimulating a wave of interest in reporting the regularity index for various retinal cell types. But just how representative the regularity index derived from a single field is, and whether there is any variation as a function of cellular density or retinal eccentricity, were issues rarely considered in many of those subsequent studies.

***Comparing the regularity index of a mosaic with that for a random point pattern***

The regularity index thus came to be regarded as an attribute of particular types of retinal cells, and while it suggested a means of comparing the orderliness between different cell types, its baseline comparison with random distributions was left undefined. Jeremy Cook (1996) first drew attention to this concern, recognizing the need to consider the size of the sampled population and the constraint imposed by the geometry of the sampled field. Cells close to the boundary (e.g. figure 1a, b) will have uncertain nearest neighbors, and the proportion of such cells in a sample will increase as the areal size of the sampled field or the density of cells therein declines, or when the aspect ratio of the sampled area changes from a square to a rectangle (Cook, 1996). The regularity index for a theoretical random distribution of points was determined to be ~1.91, providing a potential baseline that many have used to compare with real mosaics, but when actually measured from random simulations of points, was found to deviate well above or below this value with progressively smaller numbers of points in the simulated field (figure 1e). But regardless of the number of cells, the average regularity index for such random simulations approximated this theoretical value of 1.91. Note, however, that the failure to exclude points along the border (i.e. those with uncertain nearest neighbors in the random simulations) yields average regularity indices that drift off of the theoretical random value of 1.91, increasingly so as cell number declines (figure 1f).

Figure 1

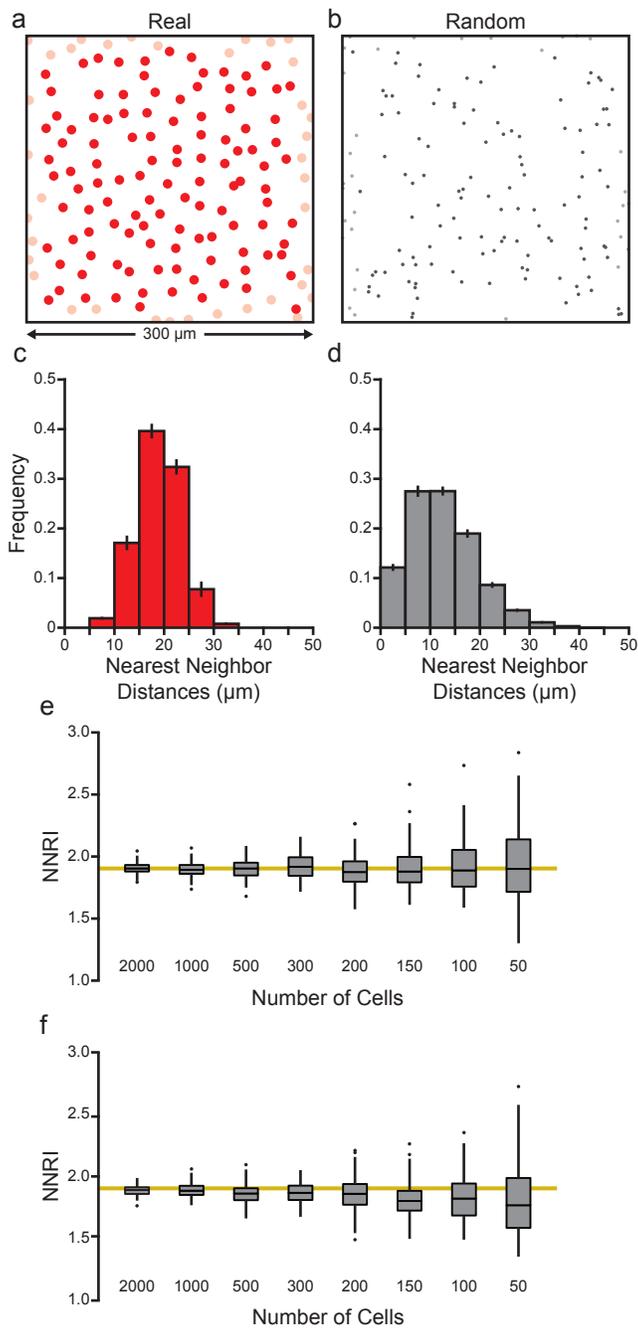

Whether or not such cells near the border are excluded, it is clear that individual random simulations at low density can veer markedly off of the theoretical value (figure 1e, f). Cook (1996) consequently generated his "Ready-Reckoner" chart identifying the critical values of the regularity index (termed the "Conformity Ratio" therein[2]) for a real sampled field as a function of the number of cells, at different levels of probability. So equipped, researchers were now able to proclaim that an analyzed field of cells was or was not significantly more orderly than a random distribution of points of the same density (e.g. Allison et al., 2010; Fite et al., 1999; Jang et al., 2011; Kolb et al., 2002; Marc, 1999b; Stenkamp et al., 2001; Yeo et al., 2009).

---

[2] The appealing features of this terminology, along with the limitations of "regularity index", are addressed in Cook (1996). But as the latter term has gained widespread usage, we adopt it herein.

***The benefit of randomly simulating distributions of cells rather than points***

Retinal mosaics are commonly confined to specific strata within the different cellular layers of the retina, and rarely are these cells observed to overlie one another. Each cell type has a characteristic somal size, and that size constrains potential proximity between two like-type neighbors. Wässle and Riemann (1978) recognized this concern as they also studied the regularity index for cone distributions in the cat and monkey retina at two different eccentricities (3 versus 13 mm), where cone density varies extensively (~10,400 versus ~3,600 cells/mm$^2$, respectively). For both species, the more central samples generated higher regularity indices (monkey: 10.9; cat: 6.5) than in the periphery (monkey: 7.0; cat: 5.1). But to assess how these regularity indices compare with random distributions (at least for the higher density samples), they simulated random distributions of cones, constraining their positioning by the average size of the cone inner segment, being ~5 µm. So constrained, an array of 209 randomly distributed elements at a comparable density of 10,000 cells/mm$^2$ yielded a regularity index of 4.4, well in excess of the thresholds defined in Cook's (1996) Ready-Reckoner chart at all four levels of significance. The real cone arrays in both cat and monkey still exceed this, but knowing this value qualifies the magnitude of those former values. Sampling the cone photoreceptor mosaic from the dorsal mouse retina, for instance, where cone density approaches this same value of 10,000 cells/mm$^2$, Fei (2003) reported a regularity index of 3.6, well below the value for the random constrained simulation above. Doubtless the size of a cone inner segment is smaller in the mouse, but absent a simulation of cones constrained by that size, we cannot judge whether the regularity of the cone mosaic in the mouse exceeds that for such a random simulation.

This constraining effect of soma size on the regularity index of random distributions, particularly as the density of simulated cells increases, can be demonstrated directly by systematically varying cellular density in simulated random retinal mosaics (from 1,000 to 5,000 cells/mm$^2$) at three different physical sizes (soma diameters of 5.0 µm, 7.5 µm and 10.0 µm) (e.g. figure 2a-f). While the smaller simulated diameter exhibited only a modest effect upon the regularity index at the five different densities examined, increasing from ~2 to ~3 as a function of increasing density, the larger simulated diameters had more profound effects, with the largest yielding an increase in the regularity index from ~3 to ~9 as density increased (figure 2g) (Keeley et al., 2017b).

Real mosaics must prohibit proximity between neighboring cells by virtue of their physical sizes, thereby ensuring that the smallest nearest neighbor distances in a distribution can never be as small as those derived from random distributions of infinitesimally small points exhibiting complete spatial randomness (CSR). Furthermore, as those distributions increase in density, the longer nearest neighbor distances that are occasionally observed at lower density will also never be sampled. Both of these features necessarily constrain the range of possible nearest neighbor distances in simulated fields of cells, making clear that a direct comparison of a real mosaic with CSR sets a very low threshold for proclamations that a mosaic is "regular". At best, one might avow that the sampled field is more regular than a random point process (although calling it "non-random" rather than "regular" would be a more apt description), but what is to be gained by such a claim when it must necessarily be the case? A more stringent claim would be based on the assessment of just how much more regular a mosaic is relative to a random simulation when that simulation is constrained by the physical size of the cells, as Wässle and Riemann (1978) demonstrated for cone photoreceptors. The comparison with a randomly distributed collection of such dimensionless points is, we suggest, largely pointless.

Figure 2

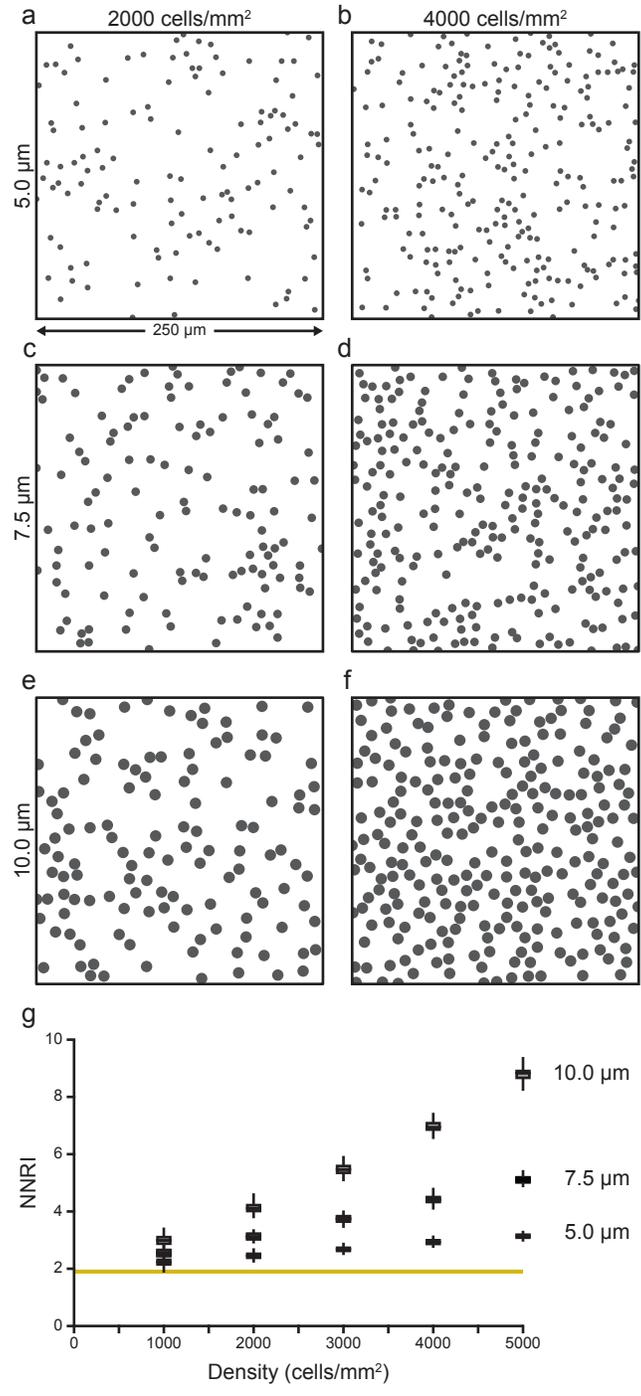

*The mosaic of AII amacrine cells—a regular retinal mosaic?*

The mosaic of AII amacrine cells, being the densest of retinal amacrine cells, illustrates this issue well. The nearest neighbor analysis of AII amacrine cells, in all species for which it has been reported, has led to the conclusion that it is a regular retinal mosaic, evidenced by regularity indices ranging from 2.7 to 5.0, depending upon the species (Gaillard et al., 2014; Jeon et al., 2007; Mills and Massey, 1991; Perez de Sevilla Müller et al., 2017; Vaney, 1985; Vaney et al., 1991; Wässle et al., 1995; Wässle et al., 1993). Yet published distributions of AII amacrine cells appear strikingly irregular, in species as diverse at mice, bats, rabbits and monkeys (figure 3a-d), exhibiting nothing like the patterning associated with other more regular retinal mosaics like the horizontal cells (compare with figure 1a). If we generate random simulations of these AII amacrine cell mosaics at equivalent densities, constrained by the average soma size gleaned from those same studies, we find the simulated fields to exhibit a comparable degree of disorder (figure 3e-h). By computing the regularity index from such random simulations of cells (in this case, generating 99 such simulations for each of these real fields), we find that the regularity indices of the real arrays fall within the range of indices derived from those random simulations, for all four species (figure 3i).

While at least some of those studies have demonstrated that the AII amacrine cell mosaic is significantly different from CSR (Diggle, 2002), characterizing it as being "regular" blurs the distinction from other orderly mosaics, such as that of the horizontal cells, which generate regularity indices far in excess of that produced by simulations of random distributions of cells (Raven and Reese, 2003). We would like to understand the features that make these mosaics qualitatively different from one another, and a simple comparison of the absolute magnitude of the regularity index doesn't begin to convey this distinction. The images (figure 3a-d) show that AII amacrine cells are occasionally positioned side by side, as occasionally are pairs of horizontal cells (e.g. figure 1a). But is the frequency of such side-by-side occurrences in those former mosaics sufficient grounds for concluding they are random? Indeed, we might go on to ask, is the distribution of *all* cells in a mosaic, as a function of distance from each cell, different from that achieved by a random distribution of cells?

Figure 3

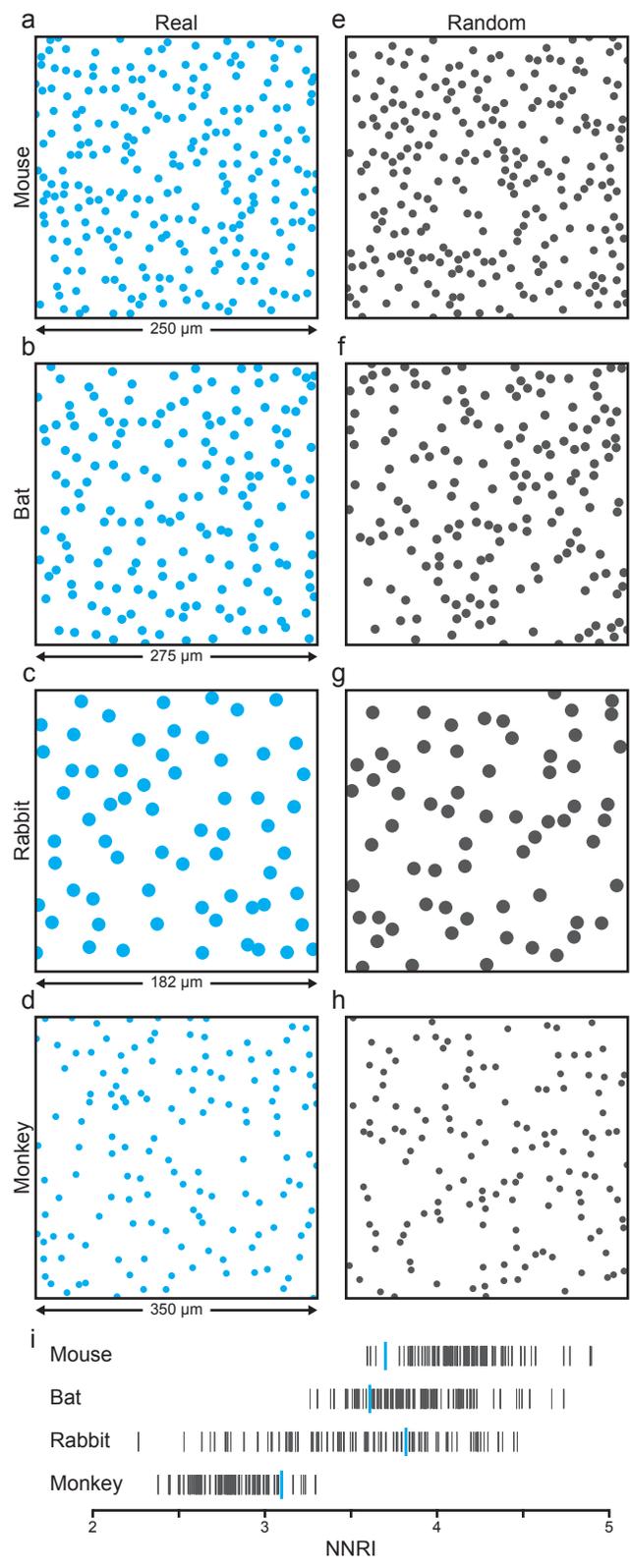

### *The density recovery profile and the effective radius*

The patterning present in regular retinal mosaics mentioned above, for example, the horizontal cells, is rarely so precise as to approximate a lattice. Such periodicity in a mosaic should reinforce itself in the spatial auto-correlogram derived from the two-dimensional somal pattern in the mosaic (being the plot of the positioning of all cells in the field relative to every other cell), yet most retinal mosaics show no evidence for it (figure 4a, b)[3]. Rather, these auto-correlograms typically show a region surrounding the origin where no cells are positioned, with a gradual increase in cellular density at further distances from the origin until average cell density is achieved. Bob Rodieck (1991) first showed the utility of deriving the density recovery profile (e.g. figure 4c) from the auto-correlogram for a field of cholinergic amacrine cells in the monkey retina, from which one may estimate the size of the region surrounding the origin where cellular density was lower than the average at further distances, termed the "effective radius"[4], often less formally called the exclusion zone. Rodieck and others were soon thereafter deriving the density recovery profile from various retinal cell types in different species (Cook and Sharma, 1995; Kouyama and Marshak, 1997; Rockhill et al., 2000; Rodieck and Marshak, 1992), showing what appeared to be a universal feature of retinal mosaics, namely, an anti-clustering tendency, by which cells in a mosaic minimize their proximity to like-type cells. Subsequently, various modeling studies confirmed that the patterning present in some real retinal mosaics could be simulated simply by constraining random distributions of cells by a minimal distance spacing rule (Galli-Resta and Novelli, 2000; Galli-Resta et al., 1999; Galli-Resta et al., 1997). Deriving the density recovery profile from the auto-correlogram of a field to show the presence of such an exclusion zone became an attractive alternative means to infer such pattern-yielding behavior in a population, and seemed tantamount (for some, at least) to demonstrating regularity.

---

[3] The spatial auto-correlation of a field of cells plots the positioning of all cells in the field relative to every other cell as a function of increasing distance, out to a specified radius. It is generated by positioning the origin of the correlogram upon the centroid of each cell successively while plotting the position of every other cell lying within the correlogram.

### *The density recovery profile is relatively immune to undersampling*

Rodieck (1991) noted, and Cook (1996) subsequently analyzed in detail, the tendency for the effective radius to remain steady despite undersampling a mosaic. Cook compared the effects of progressively undersampling a real mosaic of retinal ganglion cells upon three different statistics, the average nearest neighbor distance, the regularity index, and the effective radius, showing that while the first grew incrementally greater and the second steadily declined, the effective radius remained relatively stable until the undersampling exceeded 50% of the original population (Cook, 1996). This provided a substantial boon to those analyzing labeled mosaics of cells in which antibody penetration or reporter transgene expression was suspected to be incomplete. Provided the density recovery profile revealed a null region surrounding the origin, so, by this measure, a mosaic was said to be regular (Huberman et al., 2008; Huberman et al., 2009; Kim et al., 2008; Rivlin-Etzion et al., 2011).

---

[4] Specifically, the effective radius is defined as the radius of a cylinder with height equal to the mean density of cells and equivalent volume to the integrated region in the density recovery profile where the density falls below average cellular density.

Figure 4

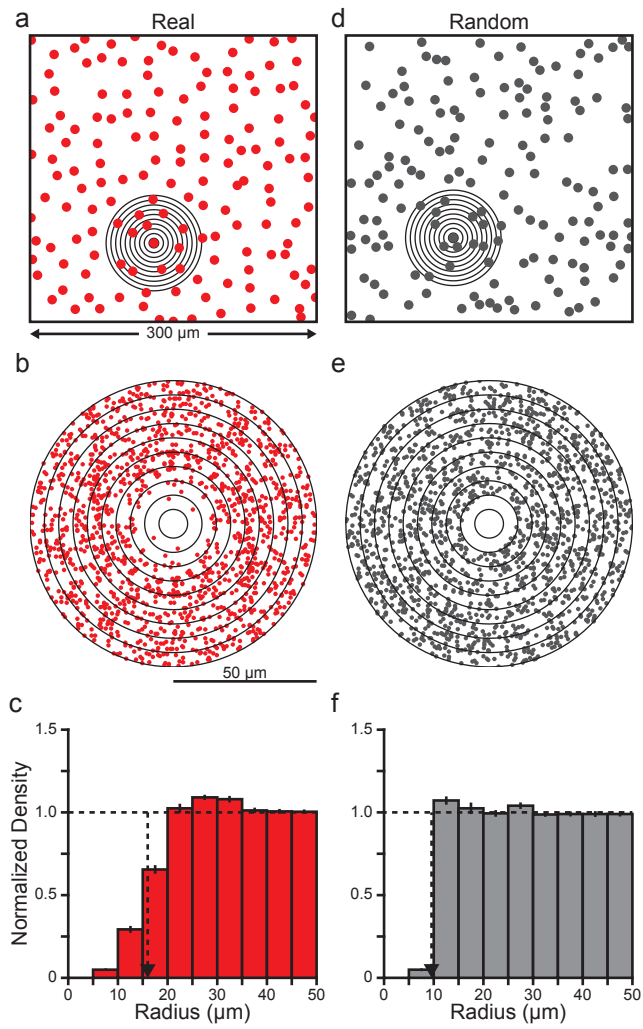

Unfortunately, it is not always straightforward to evaluate whether the details of a density recovery profile provide evidence that cells in a mosaic avoid proximity to one another, let alone whether the mosaic is regular. For instance, the high-density rod bipolar cells in the rabbit's retina yield a density recovery profile with a null region at the origin that appears no larger than the very size of the rod bipolar cells themselves (i.e. the exclusion zone is simply a consequence of soma size constraining proximity) (Rockhill et al., 2000). To consider another less clear-cut example, the far lower-density ON-OFF direction-selective retinal ganglion cells in the *DRD4-GFP* retina (Drd4 cells hereafter) exhibit a density recovery profile for a single field yielding an estimated effective radius of 25.4 µm (Huberman et al., 2009), being larger than soma size. Yet measurement of the nearest neighbor distances for this low-density population yielded a mean and standard deviation of 38.7 ± 15.7 µm (ranging from ~10-160 µm), generating a regularity index of 2.46, which, given the size of these cells, and an average density of 275 cells/mm$^2$, might lead one to ponder whether this value would exceed that derived from a random simulation constrained by soma size. Tellingly, the shape of the distribution of these nearest neighbor distances is more characteristically Poisson-like rather than Gaussian.

It is also worth noting, as Rodieck (1991) pointed out, that because the bins in the density recovery profile closer to the origin contain progressively fewer and fewer cells, they exhibit the greatest variability, so that the derivation of the effective radius is dependent upon both the density of cells in a sampled field as well as the bin size used to construct the density recovery profile. A reported effective radius for a distribution of cells can, consequently, yield an unreliable estimate of the size of the exclusion zone, particularly when dealing with low-density mosaics.

***Random simulations provide a direct assay for the effect of soma size upon the density recovery profile***

Spatial auto-correlation analysis, and its derived density recovery profile and effective radius, can, like the above consideration of the nearest neighbor analysis and regularity index, be applied to simulated random fields of cells matched for density and constrained by soma size (figure 4d). By doing so, one can directly observe the effect of soma size upon the auto-correlogram (figure 4e), the density recovery profile derived from it (figure 4f), and the associated effective radius. Regular retinal mosaics like those of the horizontal cells, for instance, show substantially larger exclusion zones in the auto-correlogram and density recovery profile relative to such random simulations (compare with figure 4a-c with d-f), yielding significantly larger effective radii as those commanded by soma size alone (Raven and Reese, 2003). But while they diverge from random simulations using these spatial statistics, it is important to appreciate that such a difference in the size of the effective radius is no indication of the magnitude of any difference in regularity between the real versus random distributions. Rather, it simply reflects the minimal spacing present within the array. Unfortunately, an insidious consequence of its increasingly widespread use has been to further the assumption that an exclusion zone is indicative of regularity in a mosaic. It is not.

***Dopaminergic amacrine cells have large exclusion zones yet are only marginally non-random***

The dopaminergic amacrine cells are one of the sparsest retinal cell types, there being an average of only ~40 cells per mm$^2$ in the C57BL/6J mouse retina. They are distributed across the retina in a pattern that appears to be random, though they are in fact non-random in their distribution (Raven et al., 2003), evidenced using nearest neighbor analysis as well as the density recovery profile (figure 5a, b). Nearest neighbor analysis shows a dearth of cells lying within 25 µm of one another, and a comparison with random simulations showed that nearest neighbor distances below 75 µm were less frequently present in the real fields (figure 5c). The density recovery profile convincingly showed this difference in minimal spacing behavior, with the recovered density in the real mosaics achieved around 100 µm, while the random simulations did so by ~40 µm (figure 5d). Because these densities were so low, requiring large bin-widths, the effective radius could not be reliably computed, but the results from this analysis suggest that dopaminergic amacrine cells also exhibit an anti-clustering tendency, avoiding proximity to one another. Despite this, these features were insufficient in yielding mosaics that were particularly regular to the eye (figure 5a), generating an average regularity index of 2.7 ± 0.1, compared to 2.0 ± 0.1 for the matched random simulations (inset in figure 5c; mean and s.e.m.), with some real fields have regularity indices falling within the range of the random simulations (see also Raven et al., 2003).

Figure 5

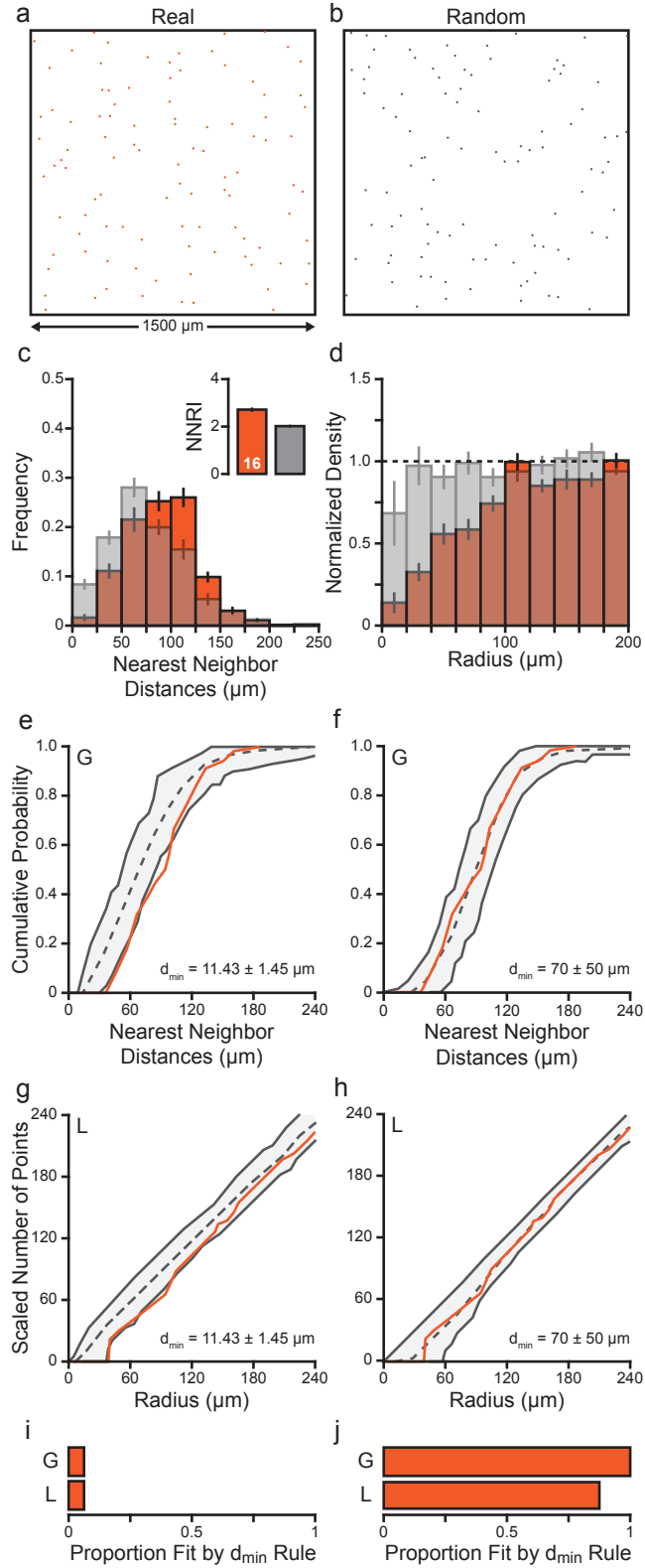

Very low cell densities should, as we saw above, also introduce greater variability in the regularity index. Where cellular density is so low, an alternative approach to comparing the regularity index or the effective radius is to plot cumulative frequency distributions for the nearest neighbor distances or for the density recovery profile. Being cumulative distributions (the G function and the L function, respectively) (Diggle, 2002; Ripley, 1976), they avoid the constraints upon the data imposed by bin size (e.g. figures 5c, d). One may then compare how well the spatial features of real fields can be simulated by minimal distance spacing rules (i.e. a spacing rule of a given mean and s.d. defining how closely a simulated cell can be randomly positioned relative to all previously positioned cells, until achieving the total number desired), by computing 99 simulations and asking whether the cumulative function for a real field would fall within the envelope of those simulations. Raven and coworkers (2003) tested the distributions of dopaminergic amacrine cells in the four densest real fields from that study, first using a minimal distance spacing rule derived from the soma size distribution of dopaminergic amacrine cells. This proved to be an ineffective fit, when comparing either the G function (the cumulative nearest neighbor frequency distribution) (figure 5e) or the L function (the cumulative density recovery profile) (figure 5g); by contrast, a larger minimal distance spacing rule of 70 ± 50 μm was successful in yielding an acceptable fit, for all four fields (figure 5g, h). (See Raven et al., 2003, for a fuller discussion of the $d_{min}$ parameters that provided acceptable fits to the real data).

Using this approach, computing these same cumulative (G and L) functions for all of the 16 sampled fields used to derive the data in figure 5c and d, along with their respective 99 simulations for each field, we confirmed that only one of the 16 fields was adequately fit by the minimal distance spacing rule based on soma size, while 14 of the 16 fields were fit by the 70 ± 50 µm spacing rule, including those fields with the lowest regularity indices (figure 5i, j).

In short, those results confirm that the dopaminergic amacrine cells avoid proximity to one another, but this simply renders them marginally non-random (e.g. figure 5a). Despite the large size of their spacing rule, their density is still far too low to yield any qualitative sense of regularity across the retina, i.e. they have a low packing intensity[5]. The packing intensity is defined as the fraction of the sample field covered by disks of fixed diameter (Eglen and Willshaw, 2002). Some of that uncovered territory within the field is due to the areas lying between densely packed circular elements; in the case of the dopaminergic cells, however, most of that unoccupied space is due to a dearth of cells (i.e. there are many locations where elements operating the same minimal distance spacing rule could be positioned, but are simply not present in the mosaic). Such a low packing intensity, coupled with a loose spacing rule, yields a patterning that is practically random (Raven et al., 2003).

***The presence of an exclusion zone does not evidence regularity in a mosaic***

In order to make even clearer the fact that the effective radius is not an index of the regularity in the patterning of an array of neurons, consider the two mosaics in figure 6, along with their associated auto-correlograms and density recovery profiles. The mosaic on the left is a random simulation constrained by a soma diameter of 10.0 ± 1.0 µm. As can be seen, the null region at the center of the auto-correlogram approximates the physical size of the simulated cells, and this is made clear in the density recovery profile, from which an effective radius was determined to be 9.9 µm.  Compare this with the mosaic on the right, composed of cells of the same physical size though at reduced density.  While not particularly regular by eye, it is immediately obvious that the elements in this mosaic exhibit a minimal spacing between themselves, and this is made apparent in the spatial auto-correlogram and in the density recovery profile.  An effective radius of 15.2 µm was derived, well in excess of the average soma size in this mosaic.  One might be inclined to conclude that the latter mosaic exhibits some degree of regularity in its patterning relative to the first mosaic.  Yet, in fact, the mosaic on the right is the very same mosaic as the original one on the left, only it has now been simulated to undergo uniform retinal expansion across the field, showing only the central region from it.

---

[5]Packing intensity can range from between 0 (for infinitesimally small points) to 0.91 (for hexagonally-arranged disks).

The point of this comparison is simply to stress the fact that two mosaics of identical patterning can have different exclusion zones.  An obvious corollary, then, is that an effective radius derived from the density recovery profile is no proxy for assessing patterning in a mosaic.  Rodieck and Marshak (1992) had previously made this very point, in their "little bang" hypothesis, observing the dimensional nature of the effective radius, and how it should be expected to vary across eccentricity as a consequence of differential retinal expansion during development, whereas indices of patterning (like the scale-invariant regularity index) should not.

***Voronoi domain analysis assesses mosaic patterning by considering the spacing between each cell with all of its immediate neighbors***

While the nearest neighbor analysis can be effective in reporting the patterning in a retinal mosaic, it has some obvious limitations, because it only considers the relationship between each cell and its closest neighbor.  Cook (1996) spelled out this concern as he pondered an analysis of ballroom dancers where the pairs of partners were at low versus high densities: the nearest neighbor analysis will return comparable regularity indices for the two conditions despite the different patterning in the two populations. Random undersampling of a real mosaic is unlikely to generate such a result, of course, but the example makes the point that very different two-dimensional distributions can have identical nearest neighbor statistics. Put simply, the nearest neighbor analysis, under the limiting condition of evenly spaced points along a line, will generate a regularity index no different from a two-dimensional collection of points having the same lattice-like periodicity. Hence, to understand the nature of the patterning in a two-dimensional array of cells, one should take into consideration the relationship between each cell and *all of its immediate neighbors*, rather than solely its *nearest* neighbor. The Voronoi tessellation of a field of cells does just this.

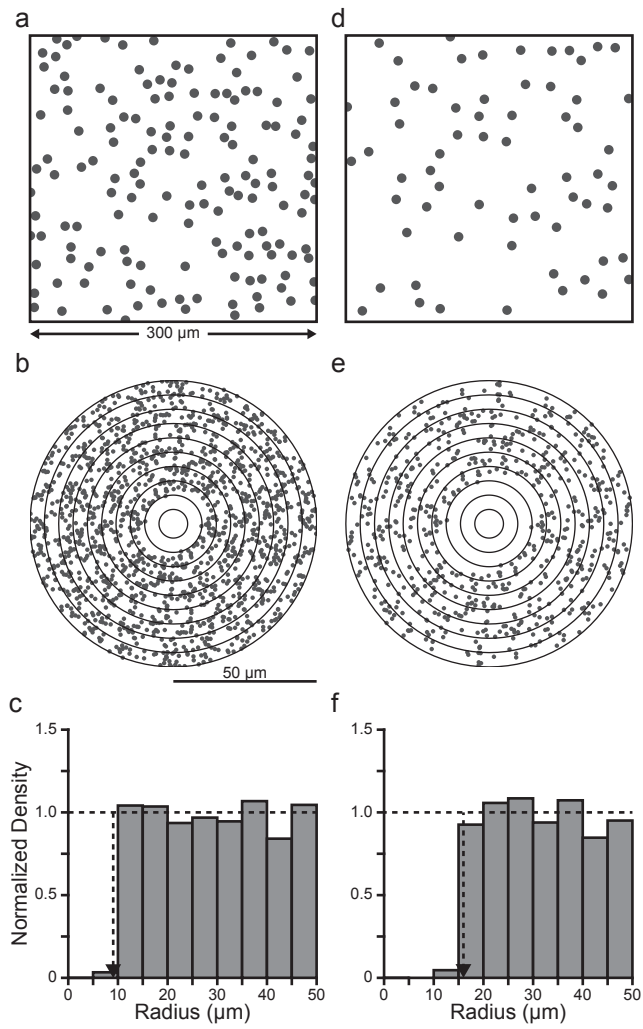

The Voronoi domain of a cell is the area in the plane of the mosaic that encloses all points closer to that cell than to any other cell. It is defined by the intersections of each of the bisectors of the Delauney segments connecting a cell to each of its immediate neighbors[6]. The Voronoi tessellation of the entire field of cells provides an immediate visual sense of the degree of regularity within the mosaic, conveyed by the variation in the size of the Voronoi areas, or "tiles", present (figure 7a). These areas, like nearest neighbor distances, can be measured and then plotted as frequency distributions; such distributions approach a Gaussian profile when reconstructed from regular retinal arrays, such as those of the cholinergic amacrine cells (shown here in figure 7a) from which a regularity index can also be computed. As with the nearest neighbor analysis and spatial auto-correlation analysis, the Voronoi domain analysis can be conducted on random simulations matched in density and constrained by soma size to show just how different a real mosaic is from random (figure 7b). Voronoi domain analysis, like the nearest neighbor analysis, is effective at reporting patterning in a retinal mosaic. But the two analyses tap into different features of such mosaics, and sometimes diverge from one another in their discriminating real from random distributions, providing insights into the factors that govern such patterning.

---

[6] Those segments interconnecting every cell to its immediate (or near) neighbors generate a triangulation of the mosaic, the Delauney tessellation of the field. The shortest of those segments is of course the nearest neighbor distance for a cell.

***The regularity index derived from Voronoi domain analysis can discriminate real from random mosaics that are not discriminated by nearest neighbor analysis***

The cholinergic amacrine cells are commonly regarded as forming a regular retinal mosaic, having been subjected to extensive analysis in various species, particularly the mouse (Galli-Resta and Ensini, 1996; Galli-Resta and Novelli, 2000; Galli-Resta et al., 2000; Galli-Resta et al., 1997; Resta et al., 2005; Sandmann et al., 1997; Vaney et al., 1981; Voigt, 1986; Whitney et al., 2008). Using the nearest neighbor analysis, for example, the population in the inner nuclear layer (INL) has been shown to be significantly more regular than random simulations constrained by soma size (figure 7e). Curiously, in some species, the mosaic in the INL appears more orderly than that in the ganglion cell layer (GCL) (Galli-Resta et al., 2000; Vaney et al., 1981; Whitney et al., 2008). In the mouse retina, when the population of cholinergic amacrine cells in the GCL is analyzed (figure 7c, d), the mosaic is found to be no more regular than random simulations constrained by soma size (figure 7f). Yet, if instead we consider the positioning of each cell relative to all of its immediate neighbors, the Voronoi tessellations of both mosaics (INL and GCL) appear more orderly than their random simulations (figure 7, compare a, c with b, d). And if the regularity index is computed from these Voronoi domain areas (VDRI), they are each confirmed to be more regular than random, while retaining the distinction of greater regularity for the INL population (Whitney et al., 2008) (figure 7g, h). Clearly, the presence of two-dimensional patterning (assessed by the regularity index) is discerned in the Voronoi analysis that is otherwise overlooked when considering only each cell's nearest neighbor.

Figure 7

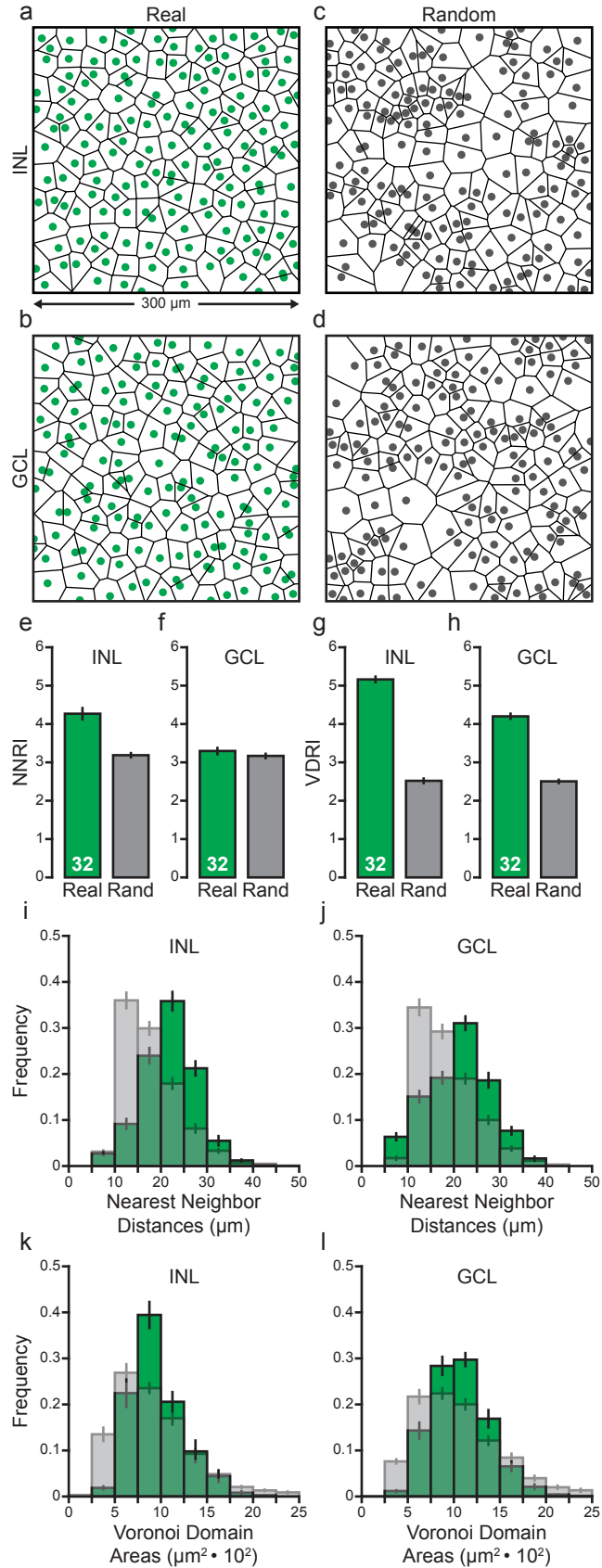

The difference in the patterning of these two mosaics arises later during development, due to a gradual deterioration in the regularity of the mosaic in the GCL. This is thought to arise from the passive lateral displacement of somata in the GCL as the retinal vasculature and optic fiber layer form (Whitney et al., 2008). Such lateral displacement of a sufficient number of cells may then yield nearest neighbor statistics that overlap those obtained for random simulations of cells while still retaining significantly different Voronoi statistics (i.e. while the smallest nearest neighbor distances found in random simulations are occasionally sampled in the GCL population, the smallest Voronoi domain areas never are).

But a closer examination of the frequency distribution of the nearest neighbor distances for the population in the GCL shows that it is not identical to that for the random simulations (figure 7j). There, the mean nearest neighbor distance is larger for the real fields, but the variance is also larger, and so the computed regularity index turns out to be identical to that generated by the random simulations, despite these differences in their spatial features (see also Galli-Resta et al., 2000). This analysis shows how reducing the analysis of nearest neighbor distances to a simple derivation of the regularity index may conceal discriminating features of the spatial patterning, presumably the very same features that permit the discrimination by a Voronoi domain-derived regularity index (figure 7h, l). In sum, a spatial analysis of patterning would be better served by tessellation-based approaches that consider each cell's relationship with all of its immediate neighbors. And while the regularity index is a handy summary statistic, the frequency distribution for either nearest neighbor distances or Voronoi domain areas itself provides a complementary and informative assessment of the spatial structure of the pattern.

***The spatial independence of retinal mosaics***

Rodieck (1991) also noted that spatial correlation analysis can be used to determine whether the positioning of cells in one mosaic has any bearing on the positioning of those in another, by computing their cross-correlogram. Cook and coworkers conducted such an analysis for different types of large retinal ganglion cells in various fish and amphibia, each individual type exhibiting a large exclusion zone in their auto-correlated condition. The resultant density recovery profile for the cross-correlated condition, by contrast, evidenced an exclusion zone no larger than soma size in every case. The cross-correlograms were essentially flat, indicating no hint of a tendency to minimize proximity between members of the two types, nor any spatial affinity between them, which would have been evidenced as an elevation in densities surrounding the origin (Cook and Sharma, 1995; Shamim et al., 1997a; Shamim et al., 1999; Shamim et al., 1997b). Rockhill and coworkers (2000) subsequently computed this analysis in the rabbit's retina for multiple pairs of mosaics, including the two populations of cholinergic amacrine cells. Essentially identical results were obtained for every pair-wise comparison examined in their study, from retinas that had been double-labeled to identify any two of six different retinal mosaics. While rare exceptions to this result have been identified (Ahnelt et al., 2000; Kouyama and Marshak, 1997), the positioning of cells in most mosaics appears to be independent of the cellular positioning within other mosaics, consistent with the general conclusions of modeling studies showing that the patterning present in a mosaic can be reproduced using a minimal distance spacing rule operating between homotypic cells that are otherwise randomly distributed.  Local spacing rules between like-type cells appear to be sufficient for recreating the patterning present in retinal mosaics (Eglen et al., 2003a; Galli-Resta, 2001).

***Two regular mosaics combined are unlikely to generate a random one***

Given such spatial independence between retinal mosaics, it is not surprising that combining the ON plus OFF populations of cholinergic amacrine cells, or of alpha or beta retinal ganglion cells, into a single composite mosaic, yields a regularity index that is far lower than the indices for their respective ON versus OFF component mosaics (Eglen and Willshaw, 2002; Wässle et al., 1981a; Wässle et al., 1981c). Whether those derived composite regularity indices (being 2.8, 2.4, 2.7, respectively) differ from random is a more complicated assessment, given that in one case, the ON versus OFF cells occupy different strata and can in principle overlie one another, whereas in the other cases, they lie in the same stratum and preclude proximity by soma size.  What seems clear is that the presence of periodicity within regular mosaics should be sufficient to ensure that a summed mosaic is not equivalent to a random simulation composed of the same total number.

Cook and Podugolnikova (2001) adopted a different approach to address this issue, by simply adding together pairs of two different retinal ganglion cell mosaics within the same retina to generate a composite mosaic, then subsequently computing the spatial auto-correlogram.  Rather than yielding a flat density recovery profile (beyond the effect of soma size) expected for a random distribution of cells, they found a profile showing reduced densities of cells surrounding the origin, though never dropping to zero (see also Rousso et al., 2016) for a more recent demonstration of the same point). Two regular mosaics then, are unlikely, when combined, to produce a distribution of cells with spatial features characteristic of a single random population.

***But two marginally non-random mosaics combined can create a more regular one***

Like the two populations of cholinergic cells in the INL and GCL, so dopaminergic amacrine cells in many carnivores are found distributed to these same two layers, rather than all being confined to the INL (Eglen et al., 2003b; Kolb et al., 1990; Oyster et al., 1985; Peichl, 1991), as in primates and rodents (Dacey, 1990; Mariani and Hokoc, 1988; Savy et al., 1989; Versaux-Botteri et al., 1984; Wulle and Schnitzer, 1989). When those two populations are separately analyzed for their mosaic regularity using Voronoi analysis, they are each *slightly* more regular than random simulations, but when combined to generate a composite mosaic, show a further increase in their regularity index (Eglen et al., 2003b), instead of exhibiting a decline as observed for the two populations of cholinergic amacrine cells (Eglen and Willshaw, 2002). And when their cross-correlogram is computed, rather than exhibiting spatial independence in their positioning, the density recovery profile yields an exclusion zone that is similar to that for the auto-correlated condition for the composite population. Results such as these suggest that cells in the two layers form a single spatial mosaic, with the less frequent cells in the GCL being misplaced from their positions within the mosaic in the INL. This interpretation is consistent with the fact that irrespective of their somal depth, they all extend their processes to arborize in the same stratum within the inner plexiform layer (Eglen et al., 2003b) (see also Weltzien et al., 2014, for a comparable demonstration for another amacrine cell type).

All of the foregoing studies should make clear that while the regularity index may be an attribute of individual retinal mosaics, it cannot, by itself, be used as a diagnostic for discriminating between a single population of like-type cells versus two different populations that share some other classifying attribute (e.g. Dyer and Cepko, 2001). Interpreting the regularity index derived from markers that under-sample populations, for instance reporter genes that may not express in every cell of a given type, yet that are also mis-expressed in some cells of another type, is particularly difficult if the latter cells cannot be reliably discriminated from the former. Still other markers such as transcription factors are increasingly appreciated for characterizing (in some cases, responsible for determining) particular cell types, but they are also often present in other cell types. For instance, the transcription factor NFIA is critical for the establishment of the population of differentiated AII amacrine cells (Keeley and Reese, 2019), yet it is also expressed in two other types of amacrine cell, as well as Müller glia and in Type 5 cone bipolar cells. Only by combining it with another marker that labels a different spectrum of cell types including the AII cells can the intersection of the two labels unambiguously reveal the complete AII amacrine cell population (Keeley and Reese, 2018b) (see also Rousso et al., 2016).

Consider, for example, the population of retinal ganglion cells expressing the transcription factor SATB2. In the mouse retina, the protein is expressed in more than one type of retinal ganglion cell, yet in the primate retina, it has been claimed to identify only a single type, one that is present at exceedingly low density (Dhande et al., 2019). Fields of cells appear conspicuously irregular, perhaps a consequence of under-sampling, or perhaps a consequence of over-sampling by inclusion of members of a second cell type; alternatively, it might truly reflect a degree of disorder in the patterning of a single cell type.  The low densities of SATB2 cells across most of the monkey's retina are in fact comparable to the density of dopaminergic amacrine cells in the mouse retina.  These SATB2 cells have a more Poisson-like distribution of nearest neighbor distances, with a tail to longer distances and with the shortest nearest neighbor distances suggestive of physical contact.  Consistent with the degree of apparent irregularity in those published somal patterns, the calculated regularity index was 2.24, said to indicate "a relatively regular mosaic" (Dhande et al., 2019).  The only comparison provided for this assertion was with random simulations of dimensionless points, which we have noted above is not much of a comparison for asserting a non-random somal distribution.  But regardless of whether this population is slightly non-random, as we saw for the dopaminergic amacrine cells in the mouse retina, or would in fact be well simulated by random distributions matched in density and constrained by soma size (as would seem likely using Voronoi domain analysis), there is no basis for concluding from such a regularity index alone that only a single type of cell is included in this sampled population (Dhande et al., 2019).  Much as we saw before that a single retinal mosaic may not necessarily generate a non-random regularity index (e.g. the AII amacrine cell mosaic), so we see here that a (potentially) non-random regularity index is hardly proof of a single retinal mosaic.

***Different spatial statistics convey distinct features in the patterning of retinal mosaics***

We have, consequently, advocated the parallel use of multiple spatial statistical analyses when attempting to discern the patterning present in retinal mosaics.  By also generating random simulations matched for density and geometry, and constrained by soma size, one may then compute this trio of analyses for both real distributions and matched random simulations, enabling a direct comparison to show how the former differ from random. And by sampling multiple fields of cells within a retina, and from multiple retinas, one can be assured that the spatial statistics typify the mosaic in question, rather than reflecting a particularly regular though atypical sampled field. We consider next our dataset derived from eight different types of retinal neurons in the mouse retina, showing considerable variability in their regularity and intercellular spacing.

***From random to regular: variation in the patterning of retinal mosaics***

Figure 8a presents micrographs of eight different types of retinal neurons in which various antibodies have been used that are widely recognized as being effective markers for those very cell types (Keeley et al., 2014a). From the digitized images of each mosaic, the X,Y coordinates of each cell are extracted, and somal patterns are generated (figure 8b). For each sampled field, a random distribution of cells is also simulated, matched to the number of cells in the real mosaic, and assigned a physical size associated with that cell type (specifically, randomly assigning a size to each simulated cell drawn from a Gaussian distribution of the same mean and standard deviation derived from the soma size distribution of the real population[7]), rejecting each cell with replacement should it overlie any portion of a previously positioned cell (figure 8c). The Voronoi domains associated to the individual cells in each of those distributions are also shown, conveying immediately the presence of patterning in the tessellation relative to what random distributions yield. Either four or eight fields were examined from each retina, sampling from a minimum of three retinas from different mice, for each cell type.

---

[7]Somal areas were traced from confocal images collected with a 40× or 60× objective for ~100-300 cells, depending upon the cell type (derived evenly from multiple retinas), and from these areas, the diameters of circles of equivalent area were determined and used to assign soma sizes in the random simulations.

Adjacent to these real versus simulated somal arrays, the three different spatial statistical analyses are presented, in which the real versus random data are conveyed in colored versus grey histograms or scatterplots, respectively. For each analysis of regularity, the frequency distributions for nearest neighbor distances (figure 8d) or Voronoi domain areas (figure 8f) is followed by a histogram showing the mean (± s.e.m.) regularity index (figure 8e & g). The two indices for each individual mosaic field are also presented as a scatterplot (figure 8h), showing the full range of the individual regularity indices from the sampled fields, and conveying a sense of their relationship to one another across the different cell types. The mean (± s.e.m.) density recovery profiles for the real fields and random simulations are then presented (figure 8i), followed by their effective radii (figure 8j), along with their "packing factors" (figure 8k). The packing factor is another scale-independent index, derived from the effective radius, conveying how closely the positioning of the cells in the field approximates a perfect hexagonal lattice (Rodieck, 1991)[8].

The first two cell types, the horizontal cells (red) and cholinergic amacrine cells, specifically, the population in the INL (green), are exemplars of regular retinal mosaics. The patterning in their arrays is conspicuous by eye, made even clearer by contrast with their density-matched random simulations (figure 8b & c). Their nearest neighbor frequency distributions convey an appreciation of their relatively Gaussian versus Poisson shapes, respectively (figure 8d). This distinction is borne out in the Voronoi domain analysis as well (figure 8f). By calculating the regularity index for each of the individual fields, we find that both statistics (derived from the nearest neighbor distances or from the Voronoi domain areas) confirm large differences between the real and random distributions (figure 8e & g), yet the magnitude of the difference is greater using the Voronoi domain analysis. For the most part, the two different indices correlate with one another, for either cell type, conveyed in their respective scatterplots (figure 8h).

---

[8]Specifically, the packing factor is the square of the ratio ER/MR, where ER is the effective radius and MR is the maximal radius possible were the same number of elements packed as a perfect hexagonal lattice. The packing factor can extend from zero, for a pattern exhibiting CSR, to 1.0, for a perfect hexagonal lattice. A perfect square lattice will have identical regularity but is not packed as efficiently as a perfect hexagonal lattice, necessarily having a lower packing factor (as well as a lower packing intensity).

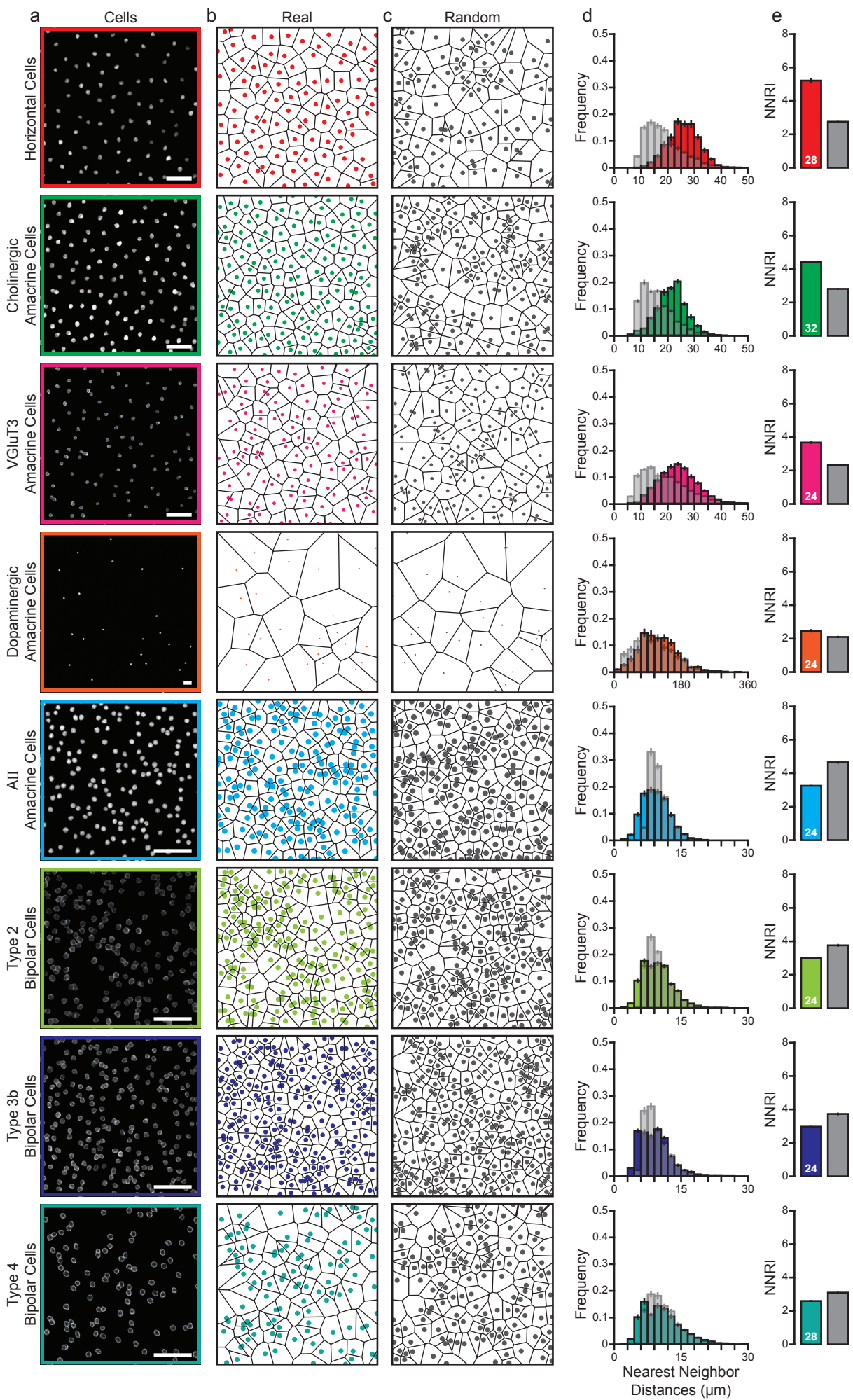

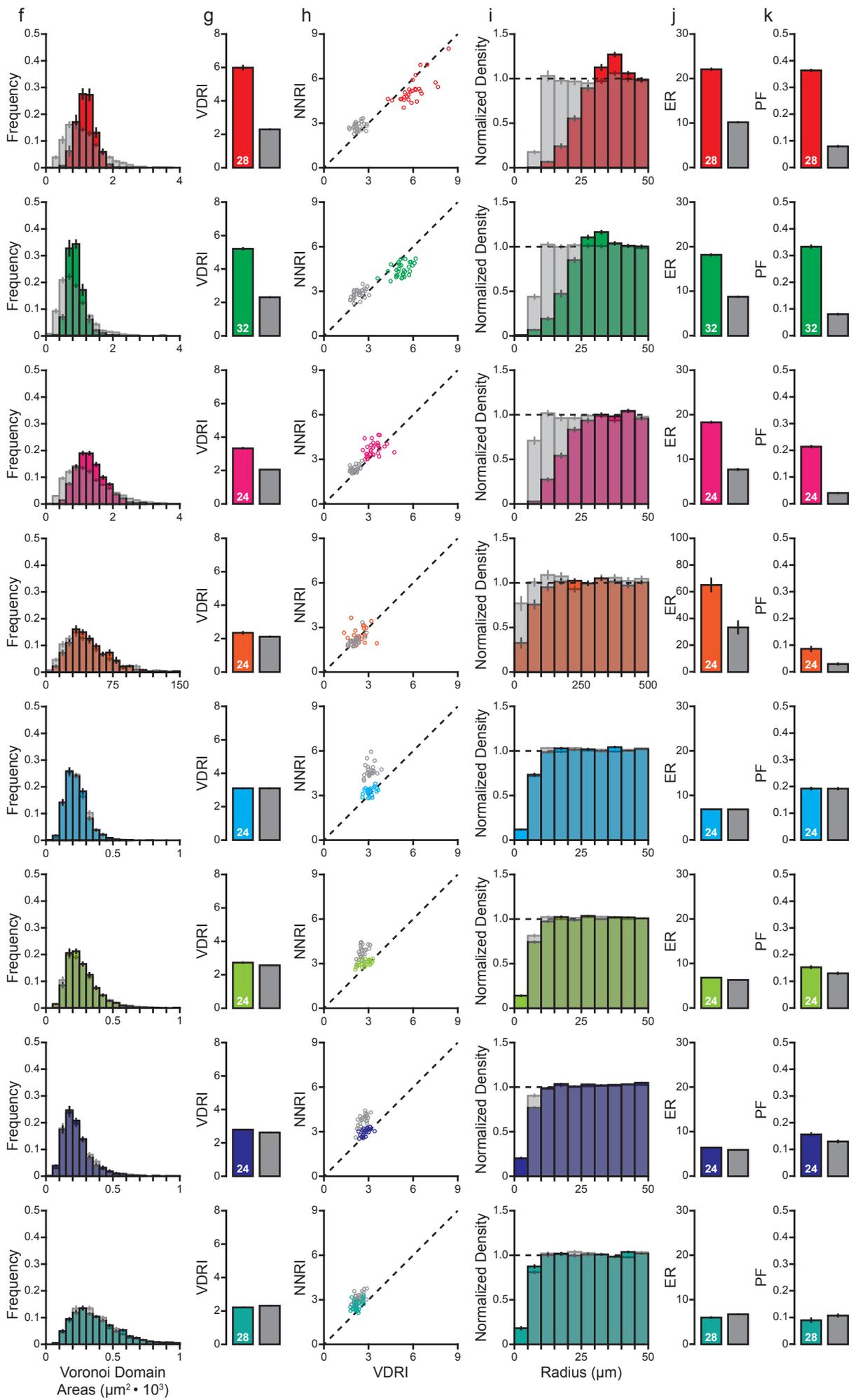

Each of these mosaics exhibits a strong tendency for cells to avoid one another. The difference between the grey and colored density recovery profiles (figure 8i) shows directly how much farther apart cells in the real array are beyond the minimal spacing imposed by soma size alone, and the effective radius for each provides a measure of this exclusion zone (figure 8j). Note that the effective radius for the random simulations returns an estimate close to the soma size used in the construction of the random simulations. Packing factor analysis, which simply expresses the size of the effective radius relative to the maximal radius permissible were this same number of cells to be packed as a perfect hexagonal lattice, also reveals a large difference from these random simulations (figure 8k). Like the two regularity indices, the packing factor is also a scale-invariant measure, but unlike them, it is not a measure of regularity per se, but rather of efficient packing (Rodieck, 1991), and is only loosely correlated to the regularity index (Reese, 2008a).

Consider next the mosaic of VGluT3 amacrine cells (magenta). Their patterning appears less orderly by comparison with either the horizontal or cholinergic amacrine cell mosaics, evidenced by the increasing breadth of their distributions (figure 8d & f), and the coincident decline in their regularity indices (figure 8e & g). The lesser regularity that is immediately apparent in the Voronoi domains (figure 8b), relative to those former cell types, is captured in the slight appearance of a tail in the Voronoi domain frequency distribution (figure 8f). Looking across these three cell types, one can observe a progressive trend for the frequency distributions (at least for the Voronoi domain areas) to become slightly more Poisson-like in shape, coincident with their progressive decline in regularity indices. While the mean effective radius of the VGluT3 cells is comparable to that for the cholinergic amacrine cells (figure 8j), their average densities are slightly lower, so that while they might exhibit the same tendency to avoid one another by this distance, there are more locations in the mosaic where cells with the same effective radius could be positioned, but are not (i.e. their packing intensity is lower). Accordingly, they also have a lower packing factor relative to either the horizontal or cholinergic amacrine cells (figure 8k), and, unlike those populations, the VGluT3 cells often have nearest neighbor regularity indices that are higher than those derived from the Voronoi domain analysis (figure 8h).

As indicated above, the dopaminergic amacrine cells (orange) are exceedingly sparse (their density in the A/J retina being less than half that found in the C57BL/6J retina in figure 5), and they lack any hint of regularity, being hard to distinguish from random distributions by eye (figure 8b & c). Yet nearest neighbor as well as Voronoi domain analyses indicate that they in fact have slightly greater average regularity indices (figure 8e & g), a consequence of the fact that they still minimize proximity to one another, borne out by the density recovery profile (figure 8i). Their very low densities render the estimated effective radii for both real and random distributions unreliable (figure 8j) (made particularly apparent by the excessive size of the estimated effective radius for the random simulations, due to the large bin widths used), even though simulations based on minimal distance spacing rules in C57BL/6J are consistent with this magnitude for the real distributions (Raven et al., 2003). Their packing factor is the lowest of all the mosaics, if not quite as low as random distributions yield, but these too should suffer from inaccurate estimation of the effective radius. Nevertheless, at the low-density end of the spectrum of all these mosaics, the analysis of dopaminergic amacrine cells serves to drive home the message that statistically non-random somal distributions can lack any obvious patterning for which the descriptor "regular" would fit.

The bottom four mosaics in figure 8, by contrast, show examples of higher density mosaics. The AII amacrine cell mosaic (cyan), as we have already considered, is noticeably irregular by eye (figure 8b & c). Interestingly, sometimes AII amacrine cells in the mouse retina are packed so close to one another such that their inter-somal distance is less than the mean somal diameter, which was derived from the mean somal *area* and used to constrain intercellular positioning in the random simulations. Somata, of course, are not perfectly round profiles, and can occasionally be compressed together, yielding some nearest neighbor distances shorter than those achieved in the random simulations (figure 8d). As a consequence, the regularity indices for the random simulations are *higher* than they are for the real arrays (figure 8e). This difference in the presence of shorter nearest neighbor distances is not apparent by eye when comparing the two somal patterns (figure 8b & c), and if one considers instead the Voronoi domain analysis, there, the distribution of domain areas, and the ensuing regularity indices, are identical (figure 8f & g; compare with d & e). The density recovery profile detects those shorter nearest neighbor distances that are absent in the random simulations (i.e. the first bin in figure 8i), yet this is such a small difference that the calculated effective radius is hardly different from that for the real distribution (figure 8j). AII amacrine cells, therefore, exhibit an exclusion zone predicted by soma size, and have a degree of patterning imposed on them solely by this constraining effect of soma size; they consequently pattern the retina no different from a random distribution of cells.

This behavior is also exhibited for three different types of retinal bipolar cells. Like the AII amacrine cells, so the mosaics of Type 2 (lime), Type 3b (blue), and Type 4 (turquoise) cone bipolar cells show a prominent disorder in their patterning, comparable to that of random distributions (figure 8b & c). Like AII cells, these cells are frequently side-by-side, and are occasionally positioned at slightly different depths yielding marginal positional overlap. Their mosaics consequently include a few very short nearest neighbor distances that are never observed in random simulations constrained by soma size (figure 8d, e & i). Yet that they are essentially random in their two-dimensional spatial statistics is made clear by their respective Voronoi analyses, where their Poisson-like distributions mimic closely those for random simulations, yielding identical regularity indices (figure 8f & g). They, like the AII cells, have effective radii that are exceedingly close to those derived for random simulations (figure 8j), indicating that their patterning upon the retina is simply the product of the spacing commanded by their physical size, nothing more.

The choice of cell types analyzed above was constrained by the availability of antibodies that selectively allow the detection of single cell types at particular depths within the retina. That the antibodies used reliably label their entire populations (i.e. the analyses are not confounded by variable undersampling), is suggested by the fact that estimates of the sizes of their entire populations, in multiple strains of mice, return an average coefficient of variation in each strain, for every cell type, around 0.04, despite different sampling procedures (Keeley et al., 2017a). Were individual retinas yielding variable proportions of a targeted cell type due to incomplete labeling of the population, we should expect greater variability in the total number of cells per retina (Keeley et al., 2014a).

***The relationship between intercellular spacing and cellular density in retinal mosaics***

To appreciate the differences in self-spacing behavior of these various cell types, consider the relationship between their spatial statistics across the variation in cellular density that is found in these sample fields. While there is some pronounced variation in the regularity indices for horizontal cells (red) at the very lowest densities (i.e. uncommonly high regularity indices are occasionally achieved when densities are so low), there is no significant correlation in either regularity index (derived from nearest neighbor distances or Voronoi domain areas) as a function of density (figure 9a & b). Likewise, the cholinergic amacrine cells (green) show no significant correlation between regularity index and density as well. In order for a mosaic to maintain regularity across variation in density, their constituent cells must modulate their average intercellular spacing, and that they do so is evidenced by the strong correlation between effective radius and density (figure 9c). Such a smooth transition in intercellular spacing in step with changes in density should yield a packing factor that also does not vary as a function of density for these cell types, as observed (figure 9d).

By contrast, in retinal mosaics wherein cells minimize proximity only by their physical size, we would expect to observe a slow increase in regularity as a function of increasing density (as shown above for the random distributions in figure 2); indeed, the three types of cone bipolar cells (lime, blue and turquoise) all appear to exhibit just such a trend across variation in density. Their effective radii, by contrast with the horizontal and cholinergic amacrine cells, should not be expected to change as a function of density (random simulations that vary in density confirm as much—Keeley et al., 2017b), and this is exactly what is obtained (figure 9c). Absent any change in effective radius in the presence of increasing cell number, their packing factor must necessarily increase, because the maximal radius must decline with that increase in density (figure 9d).

Figure 9

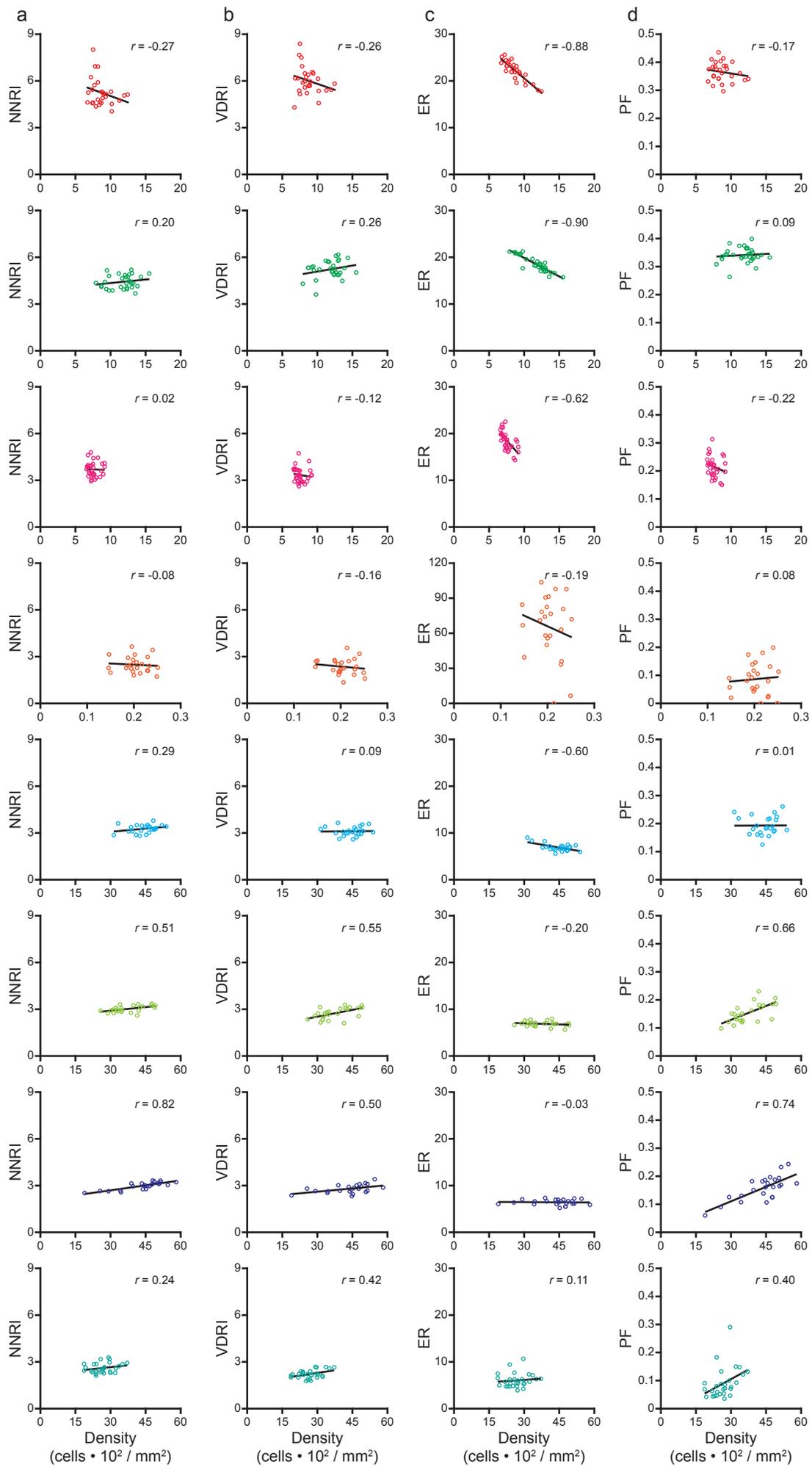

Curiously, the AII amacrine cells (cyan) present a profile slightly different from the three types of cone bipolar cells, although this appears to be due to the two sparsest fields. The trends as a function of increasing density look more like those for the three bipolar cell populations when those two fields are discounted, in particular, suggesting an invariant ER as a function of density. Regardless, whatever the cause of the slightly different trends in the AII amacrine cell mosaic (figure 9), the patterning in their mosaics cannot be discriminated from random simulations using either Voronoi domain analysis (figure 8f, g) or packing factor analysis (figure 8k), like the three bipolar cell mosaics.

The VGluT3 amacrine cells (magenta), we have already seen, are not as regular as the horizontal or cholinergic amacrine cells. Their average effective radius is comparable to that for the cholinergic amacrine cells (figure 8j), yet their somewhat lower densities must yield a lower packing intensity, and we have suggested that this yields the reduction in their regularity. If we consider their statistics as a function of density, we see that their relationship is qualitatively like that for the more regular mosaics, yet we see a strong decline in the extent of the correlation between effective radius and density for this cell type (figure 9c). As intimated above, the patterning of the VGluT3 cells may be achieved by a comparable intercellular spacing rule to that employed by the cholinergic amacrine cells, on average, yet with far greater variability in that rule (i.e. a larger s.d.) at a given density. In short, different cell types may vary in the veracity by which they execute spacing rules to minimize proximity, a notion we have already considered above for the dopaminergic amacrine cell mosaic.

The above analyses demonstrate that, for the most regular of these mosaics, the horizontal cells, there can be a degree of variability in the regularity index at the lowest densities, a variability not explained by any variation in the size of the effective radius at the same density. This result would suggest that the same spacing rule can yield a degree of variability in the ultimate patterning within a field of cells, some of which may arise by the chance positioning of small numbers of cells. Such atypically high regularity indices occasionally seen for the horizontal cell mosaic appear to be unique to this cell type, though, as neither the cholinergic nor VGluT3 amacrine cells achieve such high regularity indexes even at comparably low densities (figure 9a & b), perhaps it is only a fortuitous outcome of the largest spacing rule at such low densities (figure 9c), yielding the highest packing intensities.

***Regularity in a mosaic is a heritable trait***

The above examples of regularity in the horizontal and cholinergic amacrine cell mosaics show some degree of variability, yet their indices do not vary consistently across variation in density. Despite this, different strains of mice exhibit variation in their average regularity indices for a given cell type, which is independent of the variation in cell number across these strains, indicating that such patterning has a heritable component, being subject to genetic variants that ultimately modulate intercellular spacing (Keeley and Reese, 2014; Keeley et al., 2014b). Such variation in the regularity index across 26 recombinant inbred strains of mice mapped to a genomic locus on chromosome 11, wherein *Pttg1* is located, and which, when knocked out, reduces mosaic regularity in cholinergic amacrine cells and horizontal cells.  Results such as this, as well as the demonstration that *Megf10* and *Megf11* play a role in the development of mosaic patterning of these same two cell types (Kay et al., 2012), amply demonstrate that this histotypical property of neuronal populations can be dissected at the molecular and genetic level.

***Normalizing the regularity index enhances the detection of genomic linkage***

Interestingly, the above genomic mapping strategy supports the claim that it is the *difference* in mosaic order between real versus random distributions, rather than the absolute magnitude of the regularity index, that is the critical trait to which we should be attending (Keeley and Reese, 2014).  Because different strains of mice also vary in their absolute numbers of horizontal cells and cholinergic amacrine cells (traits that independently map to other genomic loci—Whitney et al., 2014; Whitney et al., 2011b), so then, the magnitude of the variation in regularity index across the strains will be contaminated by the space-occupying nature of each of these cell types that is progressively more constraining at higher densities.  If, instead, the regularity index of a sampled field and its random simulation are expressed as a ratio (the "regularity ratio"), the different strains of mice exhibit a change in their rank ordering of this patterning trait (figure 10a, b).  While the variation in regularity index for the horizontal cells across the strains is sufficient to map two suggestive genomic loci on chromosomes 1 and 14, the strength of genomic linkage is increased when using the regularity ratio instead (figure 10c, d) (Keeley and Reese, 2014).

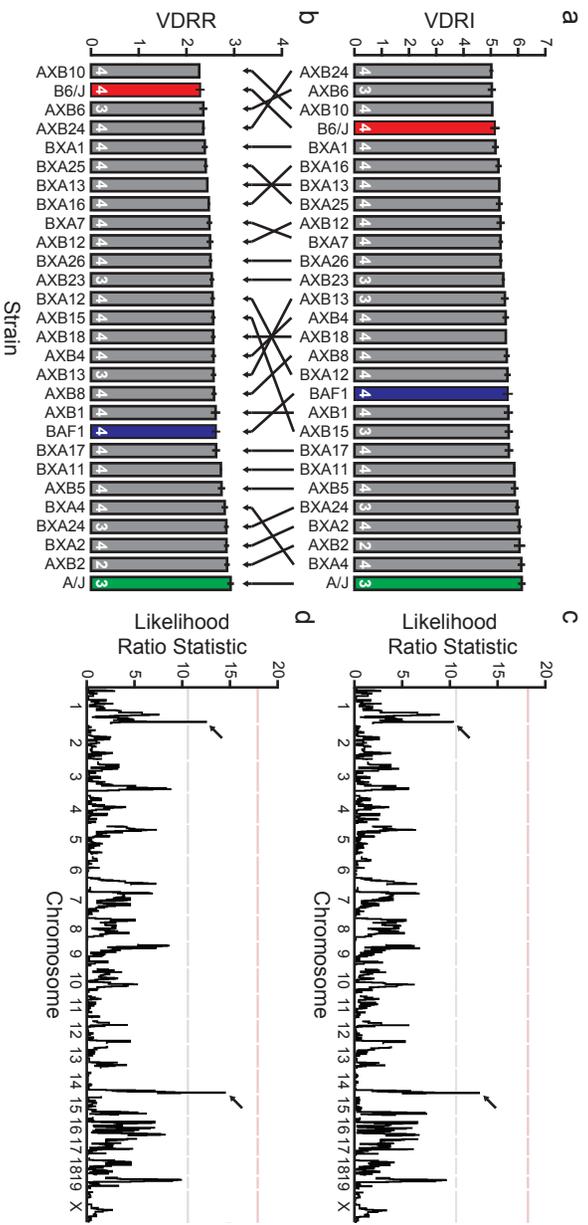

***Variation in mosaic patterning amongst retinal ganglion cell types***

The foregoing analysis shows that four out of a total of eight retinal mosaics analyzed have cellular distributions that are essentially random, while the other four mosaics have cellular distributions that are non-random though exhibiting variation in just how regular they are. As we have sampled from less than 10% of all retinal cell types, to what extent might we generalize from these results to those other cell types? In particular, no retinal ganglion cell types were included in the above analysis because we have yet to find effective markers for single types that discriminate the entire population. Yet because some well-studied populations in other species show conspicuously regular arrays (Wässle et al., 1981a; Wässle et al., 1981c), many have come to assume that all retinal ganglion cell types are assembled as regular mosaics (Seung and Sumbul, 2014; Zeng and Sanes, 2017). Each type of retinal ganglion cell is thought to extend their dendritic arbors to generate a complete and uniform coverage of the retina (Baden et al., 2016), transmitting a unique message derived from dedicated retinal circuitry at every location across the visual field. Of course, all populations of cone bipolar cells are believed to generate a uniform coverage of non-overlapping dendritic arbors (Wässle et al., 2009), and as we have already seen above, at least three of them have managed to do so without the benefit of regular spacing within their somal arrays. Likely the other types of cone bipolar cell exhibit comparable disorder while generating a dendritic tiling of the retina (but see Luo et al., 1999, for a noteworthy difference in the patterning of two types of cone bipolar cell in monkey retina). Yet there may be reasons to expect a greater variety in the organizing principles of the different types of retinal ganglion cells, as we have seen for the amacrine cells, suggested by their myriad functions (Dhande et al., 2013; Huberman et al., 2009; Kim et al., 2008;

Rousso et al., 2016; Zhang et al., 2012).  For instance, a few retinal ganglion cell types are intrinsically photosensitive, contributing to the non-image forming functions of the retina (Hughes et al., 2013), and such a luminance-detecting capacity would require neither regular somal patterning nor uniformity in process coverage.  Other retinal ganglion cell types exhibit an atypically high degree of overlap amongst their dendritic and receptive fields (Baden et al., 2016; Bae et al., 2018), while still others have uncharacteristically oriented, irregular, dendritic arbors (Huberman et al., 2009; Kim et al., 2008; Rousso et al., 2016).  What we can glean from the published literature suggests a degree of variability in the regularity of their mosaics.

    Within the mouse retina, perhaps the most convincing evidence for regularity in the mosaics of retinal ganglion cells comes from the analysis of the alpha ON sustained ($A_{ON-S}$) cell type.  These cells can be discriminated by their SMI32 immunopositive status in conjunction with dendritic arbors positioned within the inner (ON) stratum of the IPL (Bleckert et al., 2014).  While this population was not subjected to an analysis of its regularity, the density recovery profile showed an exclusion zone that increases as density declines, yielding a stable packing factor.  Other retinal ganglion cell types in the mouse retina (for example, the W3 cells and Jam-B cells) have had only their density recovery profiles reported, where the presence of an exclusion zone surrounding the origin was regarded as evidence for regular spacing, and that inferred regular spacing was then said to validate the presumption that each labeled population of cells was in fact a "natural cell type" (Kim et al., 2008; Zhang et al., 2012).  Yet images of some of these fields, by contrast with the $A_{ON-S}$ cell type, appear discernably irregular, and absent direct simulations of random distributions at the same low densities, it is difficult to assess just how different those density recovery profiles are from random at these

respective densities, let alone whether they reveal the presence of an orderly mosaic. We have, consequently, digitized the images from these various published mosaics, and generated 99 random simulations matched in density and constrained by the soma sizes derived from those published studies (figure 11), much as we described above for AII amacrine cells (figure 3), though in the present cases using both the nearest neighbor distances as well as the Voronoi tessellation to calculate each regularity index.

     The $A_{ON-S}$ cells (Bleckert et al., 2014) present a mosaic that, by eye, appears more regular than random simulations matched in density and constrained by soma size (figure 11a, f). Using either nearest neighbor analysis or Voronoi domain analysis, the regularity indexes (NNRI and VDRI) derived from the real sampled field were each substantially higher than the range of the 99 random simulated fields (figure 11k, l). This same study also reproduced mosaic fields for a second type of alpha cell, the $A_{OFF-T}$ cell, being a sparser population (figure 11b, g), but one that likewise generated regularity indexes that far exceeded the range for the 99 random simulations (figure 11k, l). By contrast, the mosaics of the Jam-B cells (figure 11c, h), the W3 cells (figure 11d, i), and the Drd4 cells (figure 11e, j), yielded regularity indexes that fell within the range of the random simulations, or in one case slightly exceeded that range on one of the two measures (Huberman et al., 2009; Kim et al., 2008; Zhang et al., 2012) (figure 11k, l). (In fact, the results for the JamB cells were associated with a p value < 0.05, but note that they hardly differ from their random simulations, unlike the $A_{ON-S}$ and $A_{OFF-T}$ cells. One would of course like to know how representative these mosaics are. Alas, only the Bleckert et al., 2014, study provided multiple sample fields of the $A_{ON-S}$ and $A_{OFF-T}$ cell types, all of them appearing comparably regular). What are we to make of this?

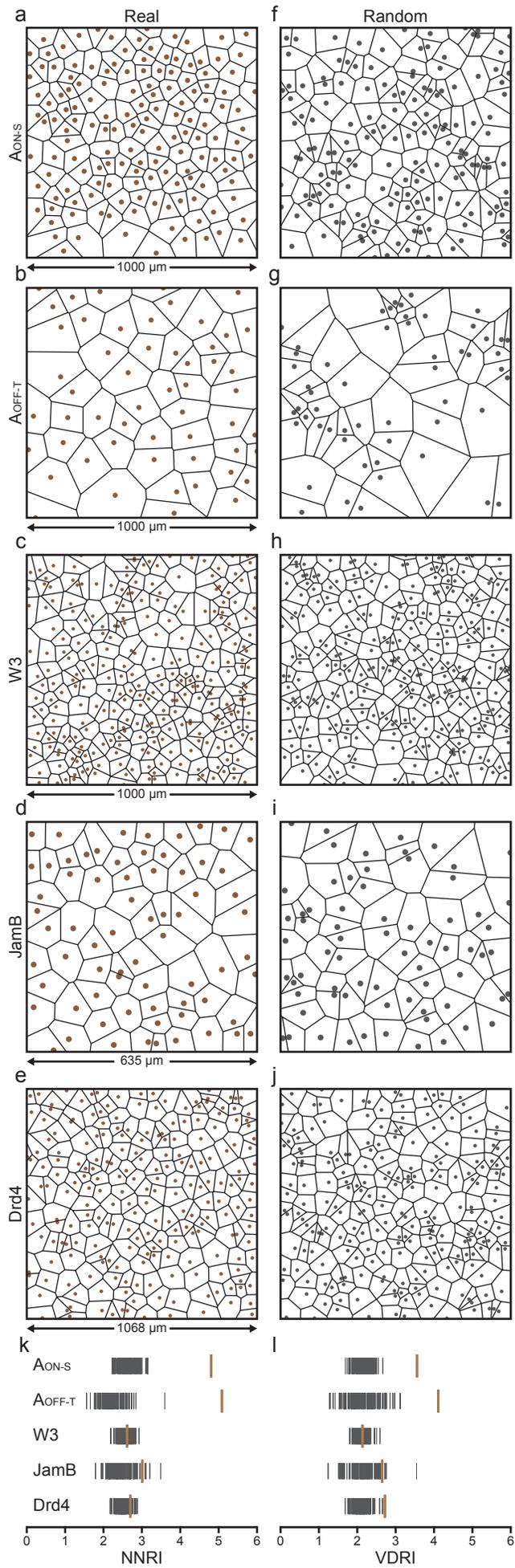

Either these three other retinal ganglion cell types have been mis-sampled in those respective studies, and might, upon a more complete or selective sampling have generated regularity indexes greater than those for matched random simulations, or these mosaics are not assembled to achieve anything like the patterning associated with the $A_{ON-S}$ and $A_{OFF-T}$ mosaics. Our own inclination is that both of these are likely to be correct: some of these studies may have mis-sampled their respective mosaics, relying as they have on variable hybridization signal or reporter gene expression, but there may also be substantial variety in their patterning, particularly when considering the variability in the organization of their dendritic arbors and associated functions. In light of the above results from other cell types, the degree of patterning (or regularity) generated by any minimal spacing behavior (evidenced by an exclusion zone) should be an empirical question rather than an *a priori* assumption about any type of retinal ganglion cell (or any retinal cell type). Many of the above studies have simply used the density recovery profile to conclude that a population of labeled cells is a single cell type based on the presence of an exclusion zone, and that its mosaic must necessarily be a regular one, particularly when it generates a regularity index exceeding that for a random distribution of infinitesimally small points (Huberman et al., 2009).

***The assembly of retinal mosaics***

The diversity in the patterning of retinal mosaics, from random to regular, raises interesting questions about their assembly, as one ultimate goal for those interested in retinal development is to understand the construction of the retinal architecture, and this includes the positioning of different types of neurons across the breadth of the retina as well as across its depth. For example, might any variation in the difference between real versus random regularity indices across species arise solely from the greater retinal expansion experienced in larger eyes (Rodieck and Marshak, 1992)? If so, these mosaics should exhibit no change in regularity or packing as a function of development. Conversely, they might show an increase in their regularity and packing at a particular developmental time-point, as do horizontal cells in the mouse retina (Raven et al., 2005b). Do differences in the relative patterning of the different ganglion cell mosaics reflect differences in their birth order? Earlier-generated types might endure a greater degree of disordering during development (like the cholinergic amacrine cells in this layer) as the optic fiber layer and retinal vasculature form (Whitney et al., 2008). And does the variation in regularity meaningfully relate to the manner by which these different cell types distribute their dendritic arbors to sample their respective afferents in the plexiform layers? The tiling behavior of the dendritic arbors of cone bipolar cells makes clear that regular patterning of somal arrays is not a requirement to generate uniform retinal coverage (Lee et al., 2011; Wässle et al., 2009); other cell types, however, exhibit variable degrees of dendritic overlap. Yet, for the moment, there is no clear relationship between the degree of mosaic regularity with homotypic regulation of dendritic growth (Reese and Keeley, 2015). These are all empirical questions, and only by addressing them directly might we come to understand the natural variation in these retinal mosaics, and the extent to which they share any organizing principles that ultimately relate to their development or function.

Kay, J.N., Chu, M.W., and Sanes, J.R. (2012). MEGF10 and MEGF11 mediate homotypic interactions required for mosaic spacing of retinal neurons. Nature *483*, 465-U117.

Keeley, P.W., Kim, J.J., Lee, S.C.S., Haverkamp, S., and Reese, B.E. (2017b). Random spatial patterning of cone bipolar cell mosaics in the mouse retina. Visual Neuroscience *34*, e002.

Keeley, P.W., and Reese, B.E. (2014). The patterning of retinal horizontal cells: normalizing the regularity index enhances the detection of genomic linkage. Frontiers in Neuroanatomy *8*, 113.

Keeley, P.W., and Reese, B.E. (2018b). DNER and NFIA are expressed by developing and mature AII amacrine cells in the mouse retina. Journal of Comparative Neurology *526*, 467-479.

Keeley, P.W., Whitney, I.E., Madsen, N.R., St. John, A.J., Borhanian, S., Leong, S.A., Williams, R.W., and Reese, B.E. (2014a). Independent genomic control of neuronal number across retinal cell types. Developmental Cell *30*, 103-109.

Keeley, P.W., Whitney, I.E., and Reese, B.E. (2017a). Genomic control of retinal cell number: Challenges, protocol, and results. In Methods in Molecular Biology: Systems Genetics, K. Schughart, and R.W. Williams, eds. (Springer), pp. 365-390.

Keeley, P.W., Zhou, C., Lu, L., Williams, R.W., Melmed, S., and Reese, B.E. (2014b). Pituitary tumor transforming gene 1 regulates the patterning of retinal mosaics. Proc Nat'l Acad Sci U S A *111*, 9295-9300.

Kim, I.-J., Zhang, Y., Yamagata, M., Meister, M., and Sanes, J.R. (2008). Molecular identification of a retinal cell type that responds to upward motion. Nature *452*, 478-482.
45

**Figure legends**

**Figure 1.** a: The mosaic of horizontal cells (red) in the mouse retina (C57BL/6J) exhibits a degree of orderliness in the positioning of its cells. b: That patterning can be contrasted with a random distribution of infinitesimally small points (CSR) at identical density (the positions of which are indicated by grey dots). Both illustrated fields contain 150 cells (or points) in a field 300 µm × 300 µm in area. c, d: The patterning in such real and simulated mosaics can be assessed by measuring the distance of each cell (or point) to its nearest neighbor, and then plotting the frequency distribution of those nearest neighbor distances. Note that elements positioned near the border (lighter shading in a and b) have uncertain nearest neighbor distances, and so are excluded from the analysis. Plotted here are the distributions for 32 real fields of horizontal cells (mean ± s.e.m.), along with 32 random simulations (each one matched in density to a real field), having densities averaging 1450 cells/mm$^2$. The two distributions differ, with that for the real data approximating a Gaussian distribution, while that for the random simulations approximates a Poisson distribution expected for a theoretical random point pattern. For each of the real fields and their random simulations, a regularity index can be computed, being the mean nearest neighbor distance divided by the standard deviation (NNRI). The average regularity index for the 32 real fields is 4.68 + 0.07, while that for the random simulations is 1.91 + 0.03. e: The regularity index for a theoretical random point pattern is ~1.91 (yellow line). Individual simulations of random point patterns can veer off of this theoretical value substantially as the number of cells in the field declines, yet they average, at any density, the theoretical value of ~1.91. Plotted here are the medians, quartiles, and outliers from 100 simulations varying in the number of cells within a sample square field. f: When points closer to the border than their nearest neighbor (e.g. lightly shaded points in b) are included in the analysis, their presence progressively distorts the average regularity index as cell number declines, taking it below the theoretical value of ~1.91 (yellow line), also evident in Cook's (1996) Ready-Reckoner chart.

**Figure 2.** Random distributions of cells that simulate increases in somal size or cell number generate increases in the regularity index. a-f: Random simulations of cells 5 (a, b), 7.5 (c, d) or 10 (e, f) μm in diameter at 2000 (a, c, e) versus 4000 (b, d, f) cells per mm$^2$. Each simulation randomly inserts a cell into the field, rejecting, with replacement, any cell that overlies a previously positioned cell, until achieving the desired density. Somal sizes are drawn from a Gaussian distribution of given mean with a standard deviation of 0.5 μm in each case. g: Variation in the nearest neighbor regularity index (NNRI) as a function of increasing density, for simulated soma sizes of 5, 7.5 and 10 μm in diameter at progressively higher densities, from 1000 to 5000 cells per mm$^2$. Plotted here are the medians and quartiles from 100 simulations at each of the densities indicated along the X axis, for the three different simulated soma sizes, in a field 250 μm × 250 μm in area. Border cells are excluded in these and all subsequent analyses. (Modified from Keeley et al., 2017).

**Figure 3.** a-d: AII amacrine cell mosaics from mouse (a), bat (b), rabbit (c) and monkey (d) are all notably irregular in their patterning. Each field shows the somal distribution of AII amacrine cells (cyan) in exemplar mosaics derived from respective publications (Jeon et al., 2007; Perez de Sevilla Müller et al., 2017; Vaney et al., 1991; Wässle et al.,

1995).  (The densities for these four fields are 3936 cells/mm$^2$ for a, from figure 1b of Perez de Sevilla Müller et al., 2017; 2221 cells/mm$^2$ for b, from figure 2a of Jeon et al., 2007; 1932 cells/mm$^2$ for c, from figure 2 of Vaney et al., 1991; and 1143 cells/mm$^2$ for d, from figure 4a of Wässle et al., 1995).  e-h: Random simulations of these same mosaics (grey), matched in area, geometry and cell density, and constrained by the soma size of AII amacrine cells reported in (or measured from) those studies are comparably irregular (somal sizes being 7.2 ± 0.4 μm for e; 8.9 ± 0.7 μm for f; 8.8 ± 0.7 μm for g; and 8.5 ± 0.7 μm for h).  The field dimensions, determined from those published images, are indicated beneath the real fields.  i: The regularity index calculated from the distribution of nearest neighbor distances (NNRI) in the four fields shown in a-d is indicated (blue), along with the regularity indices from 99 random simulations matched in density and constrained by soma size (grey), for each respective real field.  None of the regularity indices for the real fields exceeded those for the lower 95 random simulations (i.e. none were associated with a *p* value < 0.05).  The variable range of regularity indices for the 99 random simulations across the different species reflects the variation in cell number contained in the real sampled mosaics.

**Figure 4.**  a: Retinal mosaics can be assessed for any anti-clustering tendency using spatial auto-correlation analysis.  Shown is the mosaic of horizontal cells (red) from figure 1a, along with the size of the correlogram superimposed on a single soma.  b: The spatial auto-correlation of the somal distribution (i.e. the plot of every cell relative to the origin at the center of the auto-correlogram) reveals an exclusion zone, indicating a region surrounding each cell where the frequency of like-type cells is lower than at greater distances from each cell.  Such auto-correlograms generally fail to exhibit

evidence of higher-order patterning, for instance, any lattice-like periodicity, as such patterning should be amplified in the correlogram as regions of oscillating low and high density at the same periodicity.  c: The density recovery profile derived from the auto-correlogram reveals the variation in average density as a function of increasing distance from the origin.  The size of the vacant region surrounding the origin in the auto-correlogram can be estimated from the density recovery profile, termed the effective radius (vertical dashed line and arrowhead).  The analysis must either correct for cells positioned closer to the border than the radius of the correlogram, or those cells must be excluded from the analysis (Rodieck, 1991).  Plotted here are the mean (± s.e.m.) density recovery profiles from those same 32 real fields analyzed in figure 1c.  d-f: Random simulations of cells (grey; in this case, of a simulated size of 10.0 ± 0.5 µm) show the effect of soma size upon the spatial auto-correlogram and density recovery profile.  The average density recovery profile in f is derived from 32 simulated fields constrained by soma size and matched in density to the 32 real fields analyzed in c.  (The filled circles in the correlograms in b and e indicate soma positions only, not soma size).

**Figure 5.**  a, b: The dopaminergic amacrine cell mosaic (orange) in the mouse retina (C57BL/6J) is exceedingly sparse and irregular, appearing comparable to a random simulation matched in density and constrained by soma size (grey).  c: Comparison of the nearest neighbor frequency distribution (mean ± s.e.m.) derived from 16 real fields (orange) with that from 16 separately derived random simulations (each one matched for density, geometry and constrained by soma size) confirm a dearth of shorter nearest neighbor distances in the real fields.  Inset histogram shows the mean (± s.e.m.)

regularity index for these 16 fields and their random simulations. d: The density recovery profile (mean ± s.e.m.) confirms an anti-clustering tendency in the real mosaics relative to their 16 matched random simulations. Data in a-d largely replicate the original report by Raven et al. (2003), now using larger sampled fields with higher numbers of cells. e-h: Monte Carlo testing of different minimal distance spacing rules confirms that real fields are fit by spacing rules substantially larger than soma size, the latter failing to fit the data (Raven et al., 2003). Shown are the G function (the cumulative frequency distribution for nearest neighbor distances) (e, f) and the L function (the scaled cumulative density recovery profile) (g, h), comparing data from a single real field (in orange) with 99 random simulations using a $d_{min}$ based on mean soma size, of 11.43 ± 1.45 µm (e, g), versus a $d_{min}$ of 70 ± 50 µm (g, h). The average of those 99 simulations is indicated by the broken line, with the solid grey lines defining the envelope of those simulations. Informally, if the orange trace for the real data falls within the envelope of the simulations, then the model is a good fit to the data. i, j: Proportion of the 16 fields analyzed in c-e that could be fit by the two respective $d_{min}$ rules (i.e. $p < 0.95$), using the ranking procedure described in Raven et al., 2003 for assessing each field relative to its respective 99 simulations). Despite their apparently random patterning (a), dopaminergic amacrine cells minimize close proximity, rendering them non-random.

**Figure 6.** a-c: A random distribution of cells constrained by soma size (10 ± 1 µm, mean diameter ± s.d.) exhibits a null region at the center of the auto-correlogram, the size of which (discerned in the density recovery profile) approximates the soma size used in the simulation. d-f: The identical mosaic (showing only its central region in d)

has been simulated to undergo uniform expansion, consequently maintaining identical spatial relationships between the cells. The two mosaics now differ, however, in the size of the null region present at the center of their auto-correlograms, yielding a larger effective radius derived from the density recovery profile for the expanded mosaic. (The auto-correlogram and density recovery profile in e and f are derived from the full expanded field). The size of an effective radius, therefore, is not an indication of the degree of regularity in a mosaic.

**Figure 7.** a, c: Cholinergic amacrine cells in the inner nuclear layer (INL) and ganglion cell layer (GCL) differ in their regularity in the mouse retina (C57BL/6J). The fields show the Voronoi domain associated with each cell (green) in the two respective mosaics at a single locus on the retina. b, d: A comparison with random simulations constrained by soma size (grey) shows that the GCL mosaic appears more similar to the random simulation than does the INL mosaic. Note that border cells, excluded from the analysis, are defined by those that have uncertain Voronoi domains. e, f: Using nearest neighbor analysis to compute the regularity index (NNRI), the GCL mosaic is not significantly different from its random simulation. g, h: Using Voronoi domain analysis, by contrast, the regularity index (VDRI) of the GCL mosaic is significantly different from that of the random simulation, illustrating the benefit of considering all near neighbors rather than simply the nearest neighbor when assessing regularity. n = the number of fields analyzed; the number of random simulations for each comparison was identical. i, j: Frequency distributions (mean and s.e.m.) of the nearest neighbor distances for the real versus random mosaics in the INL and GCL, respectively. Note that the real versus random GCL distributions in j are clearly different, yet generating

identical mean regularity indexes (f). k, l: Frequency distributions of the Voronoi domain areas for the real versus random mosaics in the INL and GCL, respectively. Analyses were derived from 4 different retinas, sampling in the center and periphery of each retinal quadrant, for a total of 32 real sampled fields plus their respective density-matched and soma-size constrained random simulations.

**Figure 8.** Analysis of eight different types of retinal neurons in the mouse retina (A/J) show substantial variability in their patterning. Each cell type is indicated along the left margin and coded by color. a: Images of immunostained retinas, masked to reveal only the labeled somata, to accentuate their variation in patterning across cell types (see Keeley et al., 2014a, and Keeley and Reese, 2018, for unmasked examples). Calibration bar = 50 µm. b: Somal patterns derived from such immunolabeled fields, where the size of each cell is portrayed with a filled circle set to mean soma size. c: Somal patterns derived from random simulations matched in density to the fields in b, in which the simulations prohibited somal overlap. d: Average nearest neighbor frequency distribution derived from multiple fields, with data from random simulations in grey shading. e: Regularity indices calculated from the nearest neighbor distances for real fields versus random simulations. f: Average Voronoi domain area frequency distributions for real fields versus random simulations. g: Regularity indices calculated from the Voronoi domain areas for real fields versus random simulations. h: Scatterplot showing the relationship between the two regularity indices for each field. i: Average density recovery profiles derived for the real fields and random simulations. The black dashed lines indicate the normalized mean density, for each cell type. j: Effective radii computed from the density recovery profiles for the real fields versus random

simulations.  k: Packing factors computed from the effective radius for each real field versus random simulation.  Means and standard errors are plotted in all cases.  n = the number of fields analyzed; the number of random simulations for each comparison was identical.

**Figure 9.**  The different cell types vary in their behavior as a function of cellular density on four different spatial statistics.  Data are from each of the same eight cell types in figure 8e, g, j and k, now plotted for each individual field at its respective density.  a: Nearest neighbor regularity index.  b: Voronoi domain regularity index.  c: Effective radius.  d: Packing factor.  Note in particular the difference in the relationship between effective radius with density across the different cell types, and how this relates to the variation (or lack therein) of the three patterning statistics as a function of density.  (Note as well the exceedingly low effective radii and packing factors for some fields; all have unreliable effective radii and derived packing factors due to low cell density).

**Figure 10.**  a: While the regularity index does not vary systematically as a function of density *within* any strain of mice, there can be variation in the regularity index *across* different mouse strains, indicating a heritable component to this trait.  The horizontal cell mosaic of 28 different mouse strains (25 AXB/BXA recombinant inbred strains, their parental C57BL/6J and A/J strains, and their B6AF1 cross) were analyzed for their regularity index derived from their Voronoi domain areas (VDRI).  Strains are rank-ordered by regularity index.  n = the number of mice sampled per strain, with eight fields being averaged within each retina.  b: Because these strains also independently vary in horizontal cell density, the regularity index derived from matched random simulations was also determined, and the two regularity indices expressed as a ratio (real/random). As a consequence, the strains undergo a reordering by ranking this "regularity ratio" (VDRR) that expresses the magnitude of the difference from random.  c, d: Both strain distribution patterns map two quantitative trait loci (QTL), on chromosomes 1 and 14 (arrows in c and d), but this linkage between phenotype with genotype is stronger (i.e. higher likelihood ratio statistics are achieved) when normalizing the regularity index to what a density-matched random simulation would yield (d).  The horizontal lines indicate suggestive and significant thresholds defined by permutation testing of the strain data. (Modified from Keeley and Reese, 2014).

**Figure 11.**  a-e: Ganglion cell mosaics (brown) from the mouse retina derived from the published literature vary in their patterning (Bleckert et al., 2014; Huberman et al., 2009; Kim et al., 2008; Zhang et al., 2012).  Each of those studies reported that the constituent cells minimize proximity to like-type neighbors, evidenced by the presence of an exclusion zone derived from the density recovery profile, said to indicate a regular retinal mosaic, but only one of them reported the regularity index derived from nearest neighbor analysis, though comparing it to a random pattern of dimensionless points (CSR) (Huberman et al., 2009).  Included are the $A_{ON-S}$ and $A_{OFF-T}$ cells (a and b, from supplemental figures 1B and 1E of Bleckert et al., 2014), JamB cells (c, from figure 1b of Kim et al., 2008), W3 cells (d, from figure 1a of Zhang et al., 2012), and Drd4 cells (e, from supplemental figure 1A of Huberman et al., 2009).  (The densities for these five fields are 148 cells/mm$^2$ for a; 54 cells/mm$^2$ for b; 171 cells/mm$^2$ for c; 259 cells/mm$^2$ for d; and 188 cells/mm$^2$ for e).  The field dimensions, determined from those published

images, are indicated beneath the real fields. f-j: Adjacent to each real field is a random simulation, matched in density and constrained by soma size estimated by measurements from those published studies (being 20.3 ± 2.2 µm for f and inferred for g; 16.0 ± 1.1 µm for h; 15.1 ± 1.1 µm for i; and 16.9 ± 1.4 µm for j). For each real and random field, the Voronoi tessellation is indicated. But for the infrequent presence of occasional close neighbors in the random simulations, the patterning (particularly in the Voronoi tessellation) looks similar to that for the real fields for all but the $A_{ON-S}$ and $A_{OFF-T}$ cells. k, l: The regularity index derived from nearest neighbor analysis (k) and from Voronoi domain analysis (l) for each real mosaic (brown) and for the respective 99 simulations (grey). Only the $A_{ON-S}$ and $A_{OFF-T}$ cells generate large differences in regularity relative to their respective random simulations.